\documentclass[11pt,letterpaper]{article}

\usepackage[T1]{fontenc}
\usepackage{iftex}
\ifPDFTeX
  \usepackage[utf8]{inputenc}
\fi
\usepackage{lmodern}
\usepackage[margin=1in]{geometry}
\usepackage{amsmath,amssymb}
\usepackage{graphicx}
\usepackage{microtype}
\usepackage{algorithm}
\usepackage{algorithmic}
\usepackage{array,tabularx}
\usepackage[font=small,labelfont=bf]{caption}
\usepackage{placeins}
\usepackage[authoryear,round]{natbib}
\bibpunct[,]{(}{)}{;}{a}{,}{,}
\usepackage{xurl}
\usepackage[hidelinks]{hyperref}

\newcolumntype{Y}{>{\raggedright\arraybackslash}X}
\title{Fault-conditioned Seismic Image Generation using
Denoising Diffusion Probabilistic Modeling and Neural Style Transfer}

\author{
\normalsize Tolulope Agbaje$^{1}$, Nam Pham$^{2}$,
Altay Sansal$^{2}$, and Yangkang Chen$^{1}$\\[0.6em]
\small $^{1}$Bureau of Economic Geology,
The University of Texas at Austin\\
\small University Station, Box X,
Austin, Texas 78713-8924, USA\\[0.35em]
\small $^{2}$SLB, 1430 Enclave Parkway,
Suite 150, Houston, TX 77077, USA
}

\date{}

\hypersetup{
  pdftitle={Fault-conditioned Seismic Image Generation using
  Denoising Diffusion Probabilistic Modeling and Neural Style Transfer},
  pdfauthor={Tolulope Agbaje, Nam Pham, Altay Sansal, Yangkang Chen}
}

\begin{document}
\maketitle

\begin{abstract}
    Seismic data augmentation is critical for training robust deep learning (DL) models for fault detection and characterization, yet conventional approaches remain inadequate. Geometric transforms, e.g., flips or rotations, and noise injection fail to capture realistic seismic textures, while physics-based forward modeling is computationally expensive and often poorly aligned with the statistical characteristics of field data. Both approaches struggle to represent the joint distribution of seismic amplitude patterns and structural features, limiting model generalization when labeled fault examples are sparse. We present a denoising diffusion probabilistic model (DDPM) framework that learns the high-dimensional distribution of seismic images directly from field data via iterative noise injection and learned denoising. We develop two variants: an unconditional model (U-DDPM) that generates realistic post-stack seismic patches and a fault-conditioned model (FC-DDPM) that produces diverse seismic realizations that honor user-specified fault geometries encoded as binary masks. Three open-source field datasets with differing amplitude scales and processing histories are jointly standardized for training, demonstrating cross-survey generalization. Generated images from U-DDPM reproduce the dominant textural patterns, amplitude distributions, and reflector continuity observed in real seismic data. FC-DDPM maintains structural fidelity to imposed fault masks while introducing realistic variability in surrounding seismic textures, enabling one-to-many realizations from a single structural template. Frequency-wavenumber spectral analysis confirms that synthetic samples preserve the spatiotemporal bandwidth and dip content of the training distribution. The proposed framework provides a practical, controllable augmentation mechanism for expanding limited fault-labeled training sets without computationally expensive forward modeling or survey-specific parameter tuning, directly addressing a key bottleneck in DL-based seismic interpretation.  
    
\end{abstract}

\section{Introduction}
Accurate fault detection is a fundamental component of seismic structural interpretation, directly affecting assessments of hydrocarbon migration pathways, structural trapping geometries, seal integrity, and reservoir compartmentalization. Thus, the art of picking fault has received considerable attention for decades. Traditionally, seismic fault mapping relied on manual identification of reflector discontinuities, which is a subjective, labor-intensive process sensitive to interpreter experience and seismic data quality. While seismic attributes such as coherence \cite[]{chopra2005seismic, bahorich19953}, curvature \cite[]{chopra2007curvature}, and ant tracking \cite[]{pedersen2002automatic} have significantly improved discontinuity visualization, these methods remain sensitive to noise and struggle with subtle or complex fault geometries. 

With advances in technology, deep learning (DL) has been applied to automatically detect faults, overcoming limitations such as noise sensitivity, reducing bias and subjectivity, and improving adaptability and efficiency. The era of DL applications in seismic fault picking has seen the development of foundational models for automatically picking faults, including FaultNet \cite[]{zhang2019faultnet}, FaultSeg3D \cite[]{wu2019faultseg3d}, ResACEUnet \cite[]{zu2024resaceunet}, FaultVitNet \cite[]{li2025faultvitnet}, LightGEUnet \cite[]{tang2025lightgeunet}, etc. These data-driven methods have been documented to achieve impressive accuracy when sufficient training data is available. However, their performance typically depends on the quantity, quality, and diversity of labeled training data. Acquiring such labels from field data requires extensive manual interpretations, creating a bottleneck that limits the scalability and generalization of DL-based fault detection. A lack of high-quality training data could compromise the robustness of the developed model and its broader applicability to seismic data with characteristics different from those of the training data \cite[]{rolnick2017deep}. A well-suited alternative for generating training data is physics-based forward modeling, which produces synthetic seismic data with known fault labels. In addition to their simplified textural characteristics and amplitude distributions, which differ significantly from field data, generating thousands or millions of data samples may be unrealistic and computationally expensive, thereby limiting the robustness of training on a synthetic dataset and its application to field data. 

Data augmentation offers a pathway to expand the training set size. Standard data-centric augmentation techniques, such as random rotations, vertical/horizontal flips, and noise injection, can increase training diversity but may inadvertently violate geological constraints. For example, vertical flipping disrupts stratigraphic ordering, arbitrary rotations may produce geologically implausible fault orientations, and excessive smoothing attenuates the high-frequency content essential for detecting subtle fault expressions \cite[]{pham2023seismic}.  Despite a paradigm shift in the machine learning community from data-centric augmentation strategies to model-centric approaches that prioritize architectural improvements \cite[]{song2022learning,bartel2021data,pham2023seismic}, model-centric methods remain vulnerable to overfitting without adequate, high-quality, and diverse datasets.

We address this gap in realistic data augmentation through generative modeling and style transfer. Generative modeling is a broad area of machine learning that deals with models of distribution \(P(X)\) defined over data points \(X\) in some potentially high-dimensional space $\chi$ \cite[]{doersch2016tutorial}. Recent advances in deep generative models, such as generative adversarial networks (GANs) \cite[]{goodfellow2014generative} and variational autoencoders (VAEs) \cite[]{kingma2013auto}, provide a framework for learning complex, high-dimensional data distributions and have been applied to different seismic data reconstruction tasks. For example, \cite{li2020seismic} proposed a seismic labeled data expansion technique based on deep variation autoencoder. VAEs, however, primarily struggle to generate sharp, realistic images, leading to blurry outputs due to their probabilistic approach \cite[]{vivekananthan2024comparative}; a key limitation relative to GANs and diffusion models. \cite{ovcharenko2020deep} demonstrated the application of GAN for seismic data reconstruction. They noted the superiority of GANs over U-Net for interpolation tasks; however, this comes at the cost of increased uncertainty in the results and is subject to issues such as mode collapse \cite[]{vivekananthan2024comparative}. Diffusion models, such as denoising diffusion probabilistic models (DDPM) \cite[]{ho2020denoising}, offer a better alternative for data generation because they can overcome issues encountered by GANs and VAEs. However,  compared to GANs and VAEs, diffusion models are computationally intensive and require substantial training data and a large number of training timesteps. 

In this study, we propose FC-DDPM, a fault-conditioned DDPM that generates realistic post-stack seismic images, conditioned on user-specified fault geometry. Our key contribution is a two-stage training framework that decouples textural learning from structural conditioning and can be adapted to similar studies. 

\paragraph{Stage 1: Unconditional Pretraining.} We train an unconditional DDPM (U-DDPM) on unlabeled field seismic data pooled from multiple surveys. This stage learns the marginal distribution of realistic seismic textures, including amplitude statistics, frequency content, and spatial correlations, without requiring fault labels. 

\paragraph{Stage 2: Conditional Fine-tuning.} We transfer the pretrained weights and fine-tune on paired seismic-fault data using classifier-free guidance \cite[]{ho2022classifier}. To further reduce the domain gap, we apply a style transfer technique to synthetic training data, imbuing convolution modeling-derived seismic images with field-like textural and noise characteristics while preserving the fault geometry. At inference, the guidance scale parameter controls the generation process, enabling users to balance between strong adherence to fault reproduction and realistic texture diversity. This formulation also supports one-to-many generation, i.e., given a single fault mask, FC-DDPM can produce diverse seismic realizations that all honor the prescribed fault constraints while varying in textural details.

This paper is organized as follows. The Theoretical Background section \ref{subsec:DDPM} introduces denoising diffusion probabilistic models. The Methodology section presents the two-stage training framework. In the Dataset section, we describe the application of neural style transfer (NST) for domain adaptation of the synthetic data used for conditional fine-tuning. In the Model Training section, we present the U-Net architecture. We then demonstrate seismic image generation using U-DDPM and FC-DDPM. The Discussion section analyzes the role of style transfer in domain adaptation, the application of the FC-DDPM-generated dataset to DL-based seismic fault detection, and the broader implications of the presented workflow. The final section concludes.

\section{Theoretical Background}
\label{sec:section2}
\subsection{DDPM}
\label{subsec:DDPM}
In this section, we discuss the theory and mathematical formulations of DDPM based on the work of \cite{ho2020denoising}. DDPMs are latent variable models that learn data distributions through a two-process framework: a fixed forward (diffusion) process that gradually corrupts an input image with Gaussian noise, and a learned reverse (denoising) process that iteratively denoises the completely corrupted noisy image sample to recover the original data distribution \cite[]{ho2020denoising, shah2024enhancing}. 

\begin{figure}
    \centering
    \includegraphics[width=1\linewidth]{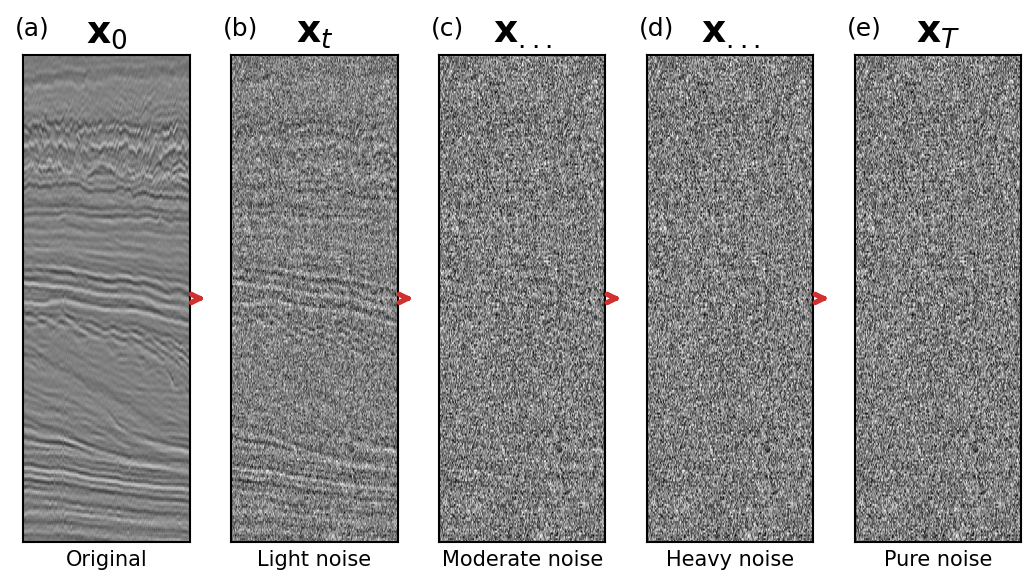}
    \caption{Illustration of the forward diffusion process applied to a seismic patch from the F3 survey. Gaussian noise is progressively added according to the variance schedule \(\bar\alpha_t\), transforming the (a) original image \(x_0\) into (d) pure noise \(x_T\) over a predefined number of timesteps, \(T\). Selected intermediate states, corresponding to (b) light noise, (c) moderate noise, and (d) heavy noise, were sampled at the 25th, 50th, and 75th percentiles of the total timesteps, \(T\). At the 75th percentile, the seismic reflectors are no longer visually distinguishable from the background noise.}
    \label{fig:ForwardDiffusionProcess}
\end{figure}

\paragraph{Forward Process:} Figure \ref{fig:ForwardDiffusionProcess} illustrates the forward diffusion process. Given a data sample \(\mathbf{x}_0 \sim q(\mathbf{x}_0)\), the forward process defines a Markov chain that progressively adds Gaussian noise over predefined \(T\) timesteps \ref{alg:forward_diffusion}. Each transition is a Gaussian perturbation with a variance schedule \(\beta_t\) that controls the magnitude of the noise injection. To ensure smooth signal-to-noise decay, we adopt the cosine noise schedule \ref{alg:cosine_schedule}. Mathematically, the forward process can be represented as follows:
\begin{equation}
\label{eq1}
q(\mathbf{x}_{1:T} \mid \mathbf{x}_0) = \prod_{t=1}^{T} q(\mathbf{x}_t \mid \mathbf{x}_{t-1}),  
\end{equation}
where \(x_1,...,x_T\) are the latent variables defined in a Markov chain and \(q(x_t|x_{t-1})\) is the Gaussian translation. The right-hand side of equation \ref{eq1}, which is the single-step Markovian transition where we add Gaussian noise, i.e., to get \(x_t\) from \(x_{t-1}\), is explicitly given as:
\begin{equation}
    q(\mathbf{x}_t \mid \mathbf{x}_{t-1}) = \mathcal{N}\!\left(\mathbf{x}_t;\ \sqrt{1-\beta_t}\,\mathbf{x}_{t-1},\ \beta_t \mathbf{I}\right).
\label{eq:eq2}
\end{equation}
A key attribute of the DDPM formulation is that we can sample \(x_t\) at an arbitrary timestep \(t\) directly from \(x_0\) without iterating through intermediate steps. The closed-form sampling can be mathematically described as:
\begin{equation}
    \mathbf{x}_t = \sqrt{\bar{\alpha}_t}\,\mathbf{x}_0 + \sqrt{1-\bar{\alpha}_t}\,\boldsymbol{\epsilon}, \qquad \boldsymbol{\epsilon} \sim \mathcal{N}(\mathbf{0},\mathbf{I}),
\label{eq:eq3}
\end{equation}
where \(\alpha_t\) = \(1-\beta_t\), and \(\beta_t\) is the pre-defined variance schedule for the forward process. \(\alpha_t\) indicates how much signal we keep and \(\beta_t\) indicates how much noise we add. As \(t\) approaches \(T\), the signal coefficient \(\sqrt{\bar\alpha_t}\) approaches \(0\) and the noise coefficient, \(\sqrt{(1-\bar\alpha_t)}\) approaches 1, ensuring \(x_T\) converges to an isotropic Gaussian. 

\begin{algorithm}[t]
\caption{Forward Diffusion Process}
\label{alg:forward_diffusion}
\begin{algorithmic}[1]
\REQUIRE Clean seismic image $\mathbf{x}_0$, timestep $t$, schedule $\bar{\alpha}_t$
\ENSURE Noisy image $\mathbf{x}_t$, sampled noise $\boldsymbol{\epsilon}$
\STATE Sample $\boldsymbol{\epsilon} \sim \mathcal{N}(\mathbf{0}, \mathbf{I})$
\STATE $\mathbf{x}_t \leftarrow \sqrt{\bar{\alpha}_t} \cdot \mathbf{x}_0 + \sqrt{1 - \bar{\alpha}_t} \cdot \boldsymbol{\epsilon}$
\RETURN $\mathbf{x}_t$, $\boldsymbol{\epsilon}$
\end{algorithmic}
\end{algorithm}

\begin{algorithm}[t]
\caption{Cosine Noise Variance Schedule}
\label{alg:cosine_schedule}
\begin{algorithmic}[1]
\REQUIRE Total timesteps $T$, offset $s = 0.008$
\ENSURE Variance schedule $\{\beta_t, \alpha_t, \bar{\alpha}_t\}_{t=1}^{T}$
\STATE $f(t) \leftarrow \cos^2\left(\frac{t/T + s}{1 + s} \cdot \frac{\pi}{2}\right)$ for $t = 0, 1, \ldots, T$
\STATE $\bar{\alpha}_t \leftarrow f(t) / f(0)$ for $t = 1, \ldots, T$
\STATE $\beta_t \leftarrow \text{clip}\left(1 - \bar{\alpha}_t / \bar{\alpha}_{t-1}, \ 10^{-4}, \ 0.999\right)$
\STATE $\alpha_t \leftarrow 1 - \beta_t$
\RETURN $\{\beta_t, \alpha_t, \bar{\alpha}_t\}_{t=1}^{T}$
\end{algorithmic}
\end{algorithm}

\paragraph{Reverse Process:} The reverse process learns to invert the forward (diffusion) process, recovering \(\mathbf{x}_0\) from \(\mathbf{x}_T \sim \mathcal{N}(\mathbf{0},\mathbf{I})\). The reverse process is also formulated in the same form as the forward process, i.e., a learned Markov chain defined by a sequence of conditional Gaussian distributions \(p_\theta(x_{t-1}|x_t)\), where the completely noisy image, \(\mathbf{x}_T\), is transformed from its Gaussian distribution, i.e., \(\mathbf{x}_T \sim \mathcal{N}(\mathbf{0}, \mathbf{I})\), to the input image \(\mathbf{x}_0\) distribution \cite[]{khader2023denoising, ho2020denoising}. The reverse process can be mathematically formulated as:
\begin{equation}
    p_\theta(\mathbf{x}_{t-1}\mid \mathbf{x}_t) = \mathcal{N}\!\left(\mathbf{x}_{t-1};\ \boldsymbol{\mu}_\theta(\mathbf{x}_t,t),\ \boldsymbol{\Sigma}_\theta(\mathbf{x}_t,t)\right),
\label{eq:eq4}
\end{equation}
where \(x_t\) is the noisy image at time step \(t\), \(x_{t-1}\) is the reconstructed less noisy image at time step \(t-1\), \(\mu_\theta(x_t,t)\) is the predicted mean, and \(\sum\nolimits_\theta(x_t,t)\) is the predicted variance.  \cite{ho2020denoising} showed that rather than predicting the mean directly, superior results are obtained by training a neural network \(\varepsilon_\theta(x_t, t)\) to predict the noise \(\varepsilon_t\) added at a given timestep \(t\) during the forward process. 

The training objective of the reverse process can be mathematically expressed as:
\begin{equation}
    L = \mathbb{E}\big[\,\lVert \boldsymbol{\epsilon} - \boldsymbol{\epsilon}_\theta(\mathbf{x}_t, t)\rVert^{2}\,\big],
\label{eq:eq5}
\end{equation}
where \(t\) is uniformly sampled from {1,...,T}. Once the network is trained, the core sampling process at each time step \(t\) is given by:
\begin{equation}
    \mathbf{x}_{t-1} = \frac{1}{\sqrt{\alpha_t}}\left(\mathbf{x}_t - \frac{1 - \alpha_t}{\sqrt{1 - \bar{\alpha}_t}}\,\boldsymbol{\epsilon}_\theta(\mathbf{x}_t, t)\right) + \sigma_t \mathbf{z}
\label{eq:6}
\end{equation}
where \(\boldsymbol{\epsilon}_\theta(\mathbf{x}_t, t)\) is the U-Net, described in detail in the U-Net subsection~\ref{subsec:UNET}, that predicts the noise \(\boldsymbol{\epsilon}\) added at timestep \(t\); \(\sigma_t\) is the standard deviation of the reverse transition (commonly set to \(\sqrt{\beta_t}\) or a learned value); and \(\mathbf{z}\) is a fresh Gaussian sample drawn at each step, i.e., \(\mathbf{z} \sim \mathcal{N}(\mathbf{0},\mathbf{I})\) (with \(\mathbf{z} = \mathbf{0}\) when \(t = 1\)).

\section{Methodology}
\subsection{Problem Formulation}
Let \(x_0 \in \mathbb{R}^{H\times W}\) denote a 2D post-stack seismic image patch and \(f \in [0,1]^{H \times W}\) a binary fault mask indicating fault locations. Our goal is to learn the conditional distribution \(p(x_0|f)\), enabling the generation of realistic seismic images that honor prescribed fault geometries.  Direct training of a conditional model on paired field seismic and corresponding fault labels is infeasible due to the scarcity of labeled field data. We therefore adopt a two-stage transfer learning approach: (1) pretrain an unconditional model on abundant unlabeled field data to capture the field texture distribution, and (2) fine-tune a conditional model on labeled style transfer-enhanced synthetic data to learn \(p(x_0|f)\) while retaining field-like characteristics. 

\subsection{Stage 1: Unconditional Pretraining on Field Data (U-DDPM)}
We pretrain an unconditional DDPM (U-DDPM) on patches extracted from multiple open-source field seismic surveys. During stage 1, the model learns the distribution of realistic seismic textures without any fault constraints. The network acquires knowledge of field-appropriate amplitude distributions, frequency content, coherent reflector patterns, and realistic noise characteristics. To stabilize optimization and obtain smoothed parameters for inference, we maintain an exponential moving average (EMA) of the network weights, which are used for all inference:
\begin{equation}
    \boldsymbol{\theta}_{\mathrm{EMA}} \leftarrow \gamma\,\boldsymbol{\theta}_{\mathrm{EMA}} + (1-\gamma)\,\boldsymbol{\theta},
\label{eq:ema}
\end{equation}
where \(\boldsymbol{\theta}\) denotes the network parameters after an optimizer update, \(\boldsymbol{\theta}_{\mathrm{EMA}}\) denotes the averaged parameters, and \(\gamma=0.999\) is the decay rate. EMA is maintained during training, and the resulting averaged parameters are held fixed during sampling.

\begin{algorithm}[t]
\caption{U-DDPM Training (Unconditional Pretraining)}
\label{alg:uddpm_training}
\begin{algorithmic}[1]
\REQUIRE Field seismic patches $\{\mathbf{x}_0^{(i)}\}_{i=1}^{N}$, schedule $\bar{\alpha}_t$
\REQUIRE U-Net $\boldsymbol{\epsilon}_\theta$ with parameters $\theta$
\REPEAT
    \STATE Sample $\mathbf{x}_0 \sim p_{\text{field}}(\mathbf{x})$
    \STATE Sample $t \sim \text{Uniform}(\{1, \ldots, T\})$
    \STATE Sample $\boldsymbol{\epsilon} \sim \mathcal{N}(\mathbf{0}, \mathbf{I})$
    \STATE $\mathbf{x}_t \leftarrow \sqrt{\bar{\alpha}_t} \cdot \mathbf{x}_0 + \sqrt{1 - \bar{\alpha}_t} \cdot \boldsymbol{\epsilon}$
    \STATE Take gradient step on $\nabla_\theta \left\| \boldsymbol{\epsilon} - \boldsymbol{\epsilon}_\theta(\mathbf{x}_t, t) \right\|^2$
\UNTIL{converged}
\RETURN Pretrained parameters $\theta$
\end{algorithmic}
\end{algorithm}

\subsection{Stage 2: Conditional Fine-tuning with Classifier-Free Guidance (CFG)}
Starting from the pretrained unconditional weights, we fine-tune a fault-conditioned model on the stylized synthetic dataset, consisting of seismic patches and their corresponding fault labels. We denote the original fault mask by \(\mathbf{f}\) and the actual conditioning input supplied to the network during training by \(\mathbf{y}\). With probability \(p_{\mathrm{uncond}}\), we set \(\mathbf{y}=\mathbf{0}\); otherwise, we set \(\mathbf{y}=\mathbf{f}\). The noisy seismic image \(\mathbf{x}_t\) is then concatenated channel-wise with \(\mathbf{y}\) to form a two-channel input. 

Because the unconditional model (U-DDPM) is pretrained with a \emph{single-channel} input (the noisy seismic image only), adapting it to the two-channel conditional setting requires modifying only the first convolutional layer. All backbone parameters, i.e., encoder, bottleneck, decoder, time-embedding, normalization, and output layers, are transferred directly from the pretrained model. The input-convolution weight is expanded from shape \((C_{\mathrm{base}}\times 1\times 3\times 3)\) to \((C_{\mathrm{base}}\times 2\times 3\times 3)\): the slice acting on the seismic channel is copied from the pretrained weights, while the slice acting on the newly added fault-mask channel is initialized to zero. With a zero-initialized mask channel, the fine-tuned model initially reproduces the unconditional predictions and ignores the mask; the conditioning signal is then acquired gradually during fine-tuning. This preserves the learned texture prior while introducing structural control in a stable manner.


To balance the adherence to fault conditioning against the preservation of realistic seismic texture, we employ classifier-free guidance method \cite[]{ho2022classifier}. During training, \(\mathbf{y}\) equals the original fault mask \(\mathbf{f}\) with probability \(1-p_{\mathrm{uncond}}\) and a zero-valued mask with probability \(p_{\mathrm{uncond}}\). At inference, a guidance scaling parameter is applied to combine the conditional and unconditional noise estimates as:

\begin{equation}
\boldsymbol{\epsilon}_\theta^{\mathrm{CFG}}
(\mathbf{x}_t,t,\mathbf{f}) =
\boldsymbol{\epsilon}_\theta(\mathbf{x}_t,t,\mathbf{0}) 
+s\Big[\boldsymbol{\epsilon}_\theta (\mathbf{x}_t,t,\mathbf{f}) -\boldsymbol{\epsilon}_\theta (\mathbf{x}_t,t,\mathbf{0})\Big].
\label{eq:cfg}
\end{equation}

where \(s\) controls the guidance strength. The conditional and unconditional noise predictions are obtained from the same fine-tuned network using the fault mask \(\mathbf{f}\) and a zero-valued mask \(\mathbf{0}\), respectively. When \(s=0\), the expression gives the unconditional prediction; when \(s=1\), it gives the conditional
prediction without additional guidance. Values \(s>1\) amplify the difference between the conditional and unconditional predictions. During sampling, both predictions are evaluated using the fixed EMA parameters defined in equation~\ref{eq:ema}.

For fault-conditioned sampling, we use the guided noise prediction in equation~\ref{eq:cfg} in the reverse-diffusion update of equation~\ref{eq:6}, yielding equation~\ref{eq:9} below.

\begin{equation}
\mathbf{x}_{t-1} =
\frac{1}{\sqrt{\alpha_t}}
\left(
\mathbf{x}_t -
\frac{1 - \alpha_t}{\sqrt{1 - \bar{\alpha}_t}}
\boldsymbol{\epsilon}_\theta(\mathbf{x}_t, t, \mathbf{f})
\right)
+ \sigma_t \mathbf{z},
\label{eq:9}
\end{equation}
where \( z \sim \mathcal{N}(0, I) \), and \(\sigma_t\) controls stochasticity during sampling, causing the generated seismic realizations not to be arbitrary but statistically consistent with both the learned seismic data distribution and the prescribed fault geometry configuration. This formulation enables FC-DDPM to generate diverse yet geologically plausible seismic images conditioned on structural priors, providing a principled mechanism for controlled subsurface image synthesis and data augmentation in DL-based fault-detection tasks. 

\begin{algorithm}[t]
\caption{FC-DDPM Training (Conditional Fine-tuning)}
\label{alg:fcddpm_training}
\begin{algorithmic}[1]
\REQUIRE Stylized seismic patches $\{\mathbf{x}_0^{(i)}\}$, fault masks $\{\mathbf{f}^{(i)}\}$
\REQUIRE Pretrained parameters $\theta$, dropout probability $p_{\text{uncond}}$
\REQUIRE U-Net $\boldsymbol{\epsilon}_\theta$ with conditioning input
\STATE Initialize $\boldsymbol{\epsilon}_\theta$ with pretrained $\theta$
\REPEAT
    \STATE Sample $(\mathbf{x}_0, \mathbf{f}) \sim p_{\text{stylized}}(\mathbf{x}, \mathbf{f})$
    \STATE Sample $t \sim \text{Uniform}(\{1, \ldots, T\})$
    \STATE Sample $\boldsymbol{\epsilon} \sim \mathcal{N}(\mathbf{0}, \mathbf{I})$
    \STATE $\mathbf{x}_t \leftarrow \sqrt{\bar{\alpha}_t} \cdot \mathbf{x}_0 + \sqrt{1 - \bar{\alpha}_t} \cdot \boldsymbol{\epsilon}$
    \IF{$\text{rand}() < p_{\text{uncond}}$}
        \STATE $\mathbf{y} \leftarrow \mathbf{0}$ \COMMENT{Drop the fault condition}
    \ELSE
        \STATE $\mathbf{y} \leftarrow \mathbf{f}$ 
        \COMMENT{Retain the fault condition}
    \ENDIF
    \STATE Take gradient step on $\nabla_\theta \left\| \boldsymbol{\epsilon} - \boldsymbol{\epsilon}_\theta(\mathbf{x}_t, t, \mathbf{y}) \right\|^2$
\UNTIL{converged}
\RETURN Fine-tuned parameters $\theta$
\end{algorithmic}
\end{algorithm}

\subsection{NST for Domain Adaptation}
We used three open-source datasets, namely F3, Teapot, and Kerry, for U-DDPM training (Figure \ref{fig: Field Data - F3_Kerry_Teapot}).  
The F3 dataset has a dimension of \(651 \times 951 \times 463\) corresponding to the inline, crossline, and time components. Kerry, on the other hand has a dimension of \(287 \times 735 \times 1252\) while Teapot has a dimension of \(345 \times 188 \times 1501\). We applied a local normalization technique (Equations \ref{eq:eq10}--\ref{eq:eq14}) to each of the three seismic datasets, normalizing the range to -1 to 1. The local normalization technique ensures that all data are appropriately scaled because the three seismic datasets used have different absolute amplitude scales, acquisition gains, and processing histories. Global scaling may cause high-amplitude surveys to dominate training and low-amplitude surveys to collapse toward zero, leading to unstable diffusion training. Before scaling, we had patched each of the seismic volume into several 2D patches with dimensions \((H \times W = 256 \times 256)\) and a stride dimension of \(128 \times 128\). We then normalize locally per patch, preserving local contrast. The mathematical formulation of the normalization steps is described below:

\begin{figure}
    \centering
    \includegraphics[width=1\linewidth]{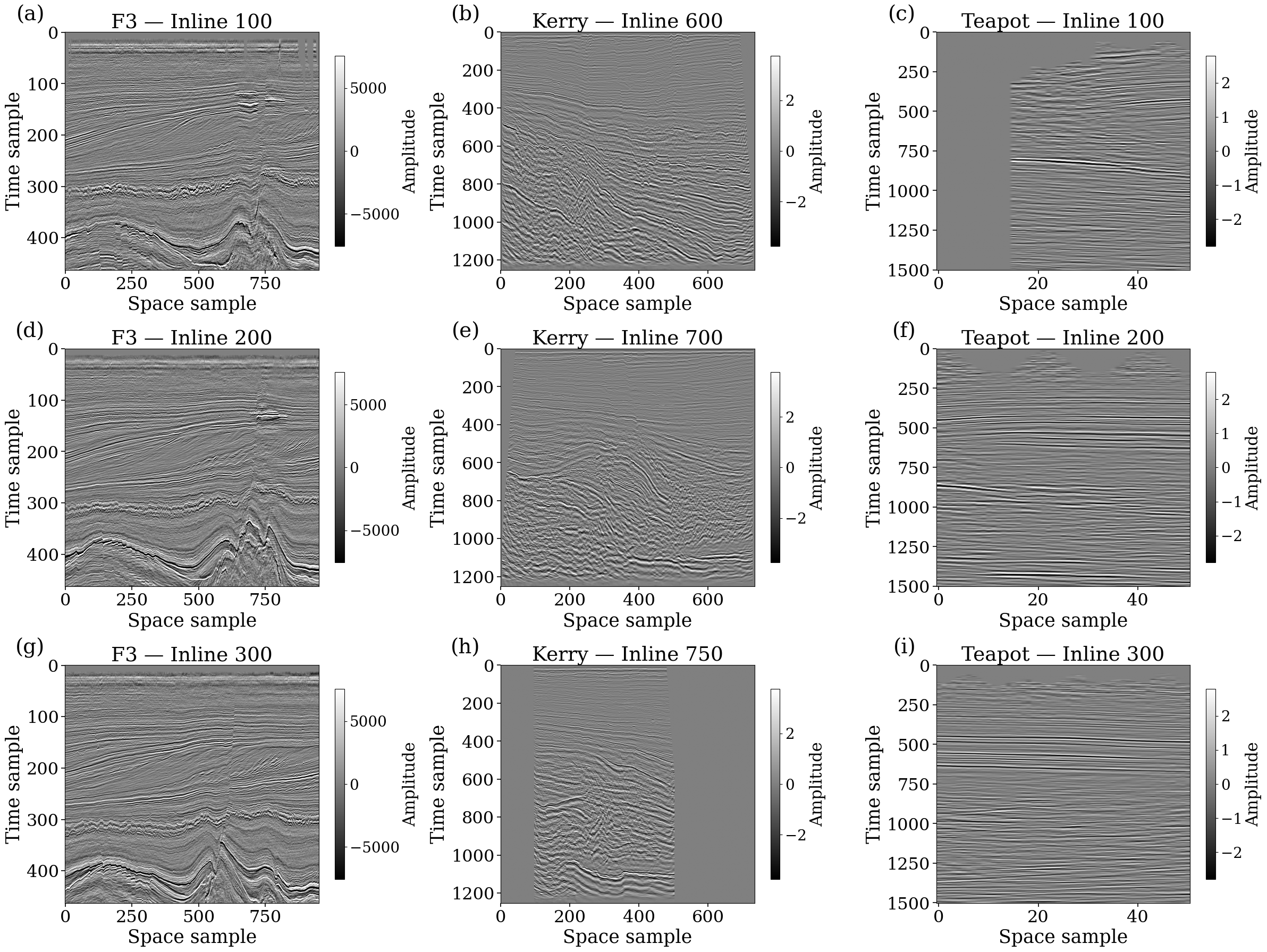}
    \caption{Representative seismic sections from open-source datasets used in this study. Left (a,d,g): inline sections from the F3 survey, which serves as the most representative field dataset for unconditional pretraining. Center (b,e,h): section from the Kerry 3D survey. Right (c,f,i): section from the Teapot Dome survey.}
    \label{fig: Field Data - F3_Kerry_Teapot}
\end{figure}

Let $A \in \mathcal{R}^{N\times M\times T}$ denote a 3D post-stack seismic amplitude volume, where \(N\) is the inline index, \(M\) is the crossline index, and \(T\) represents the time samples. We extract 2D seismic sections and further extract local patches. Let a single 2D patch be denoted as:
\begin{equation}
    P = {p_{ij} \in \mathcal{R}^{H \times W}},
\label{eq:eq10}
\end{equation}
where \(H\) and \(W\) is the patch size, i.e., \(256 \times 256\) (Figure 3). For a given patch, we define a robust amplitude scale using a high quantile of absolute amplitudes:
\begin{equation}
    s(P) = Q_q({|p_ij|_{ij}}),
\label{eq:eq11}
\end{equation}
where \(Q_q(\bullet)\) is the \(q-th\) quantile (e.g., \(q = 0.99\)). To ensure numerical stability, we define 
\begin{equation}
    \bar s(P) = max(s(P), \varepsilon),
\label{eq:eq12}
\end{equation}
with \(\varepsilon \ll 1, e.g., \varepsilon = 10^{-6}\). Each pixel in the patch is then normalized as :
\begin{equation}
    \bar p_{ij} = \frac{p_{ij}}{\bar s(P)}.
\label{eq:eq13}
\end{equation}
This helps to map the patch into an approximately unit-amplitude range while preserving relative amplitude contrasts within the patch. Finally, we applied an enforced lower and upper limit to symmetrically clip the normalized range to -1 and 1. This helps to ensure that \(p_{ij} \in [-1,1]\). The final normalized patch is illustrated below as:
\begin{equation}
    \mathcal{N}(P) = \text{clip}(\frac{P}{\max(Q_q(|P|), \varepsilon)}, -1,1).
\label{eq:eq14}
\end{equation}

After patching and scaling, we filtered out patches with zero values to avoid degrading the quality of images fed into the model. The filtered patches, with a dimension of \((27634 \times 256 \times 256)\), were then used for model training.  

\begin{table}[t]
\centering
\caption{Datasets and their roles in this study, distinguishing pretraining, style reference, conditional fine-tuning, and evaluation.}
\label{tab:dataset_usage}
\begin{tabularx}{\linewidth}{@{}lYY@{}}
\hline
\textbf{Dataset} & \textbf{Role} & \textbf{Stage / split} \\
\hline
F3                 & Pretraining; NST style reference; field test & U-DDPM; NST; field evaluation \\
Kerry              & Pretraining (texture diversity)              & U-DDPM \\
Teapot Dome        & Pretraining (texture diversity)              & U-DDPM \\
FaultSeg3D (synth.)& Conditional fine-tuning; labeled train/val   & FC-DDPM; downstream train/val \\
\hline
\end{tabularx}
\end{table}

\begin{figure}[tbp]
    \centering
    \includegraphics[width=1.0\linewidth]{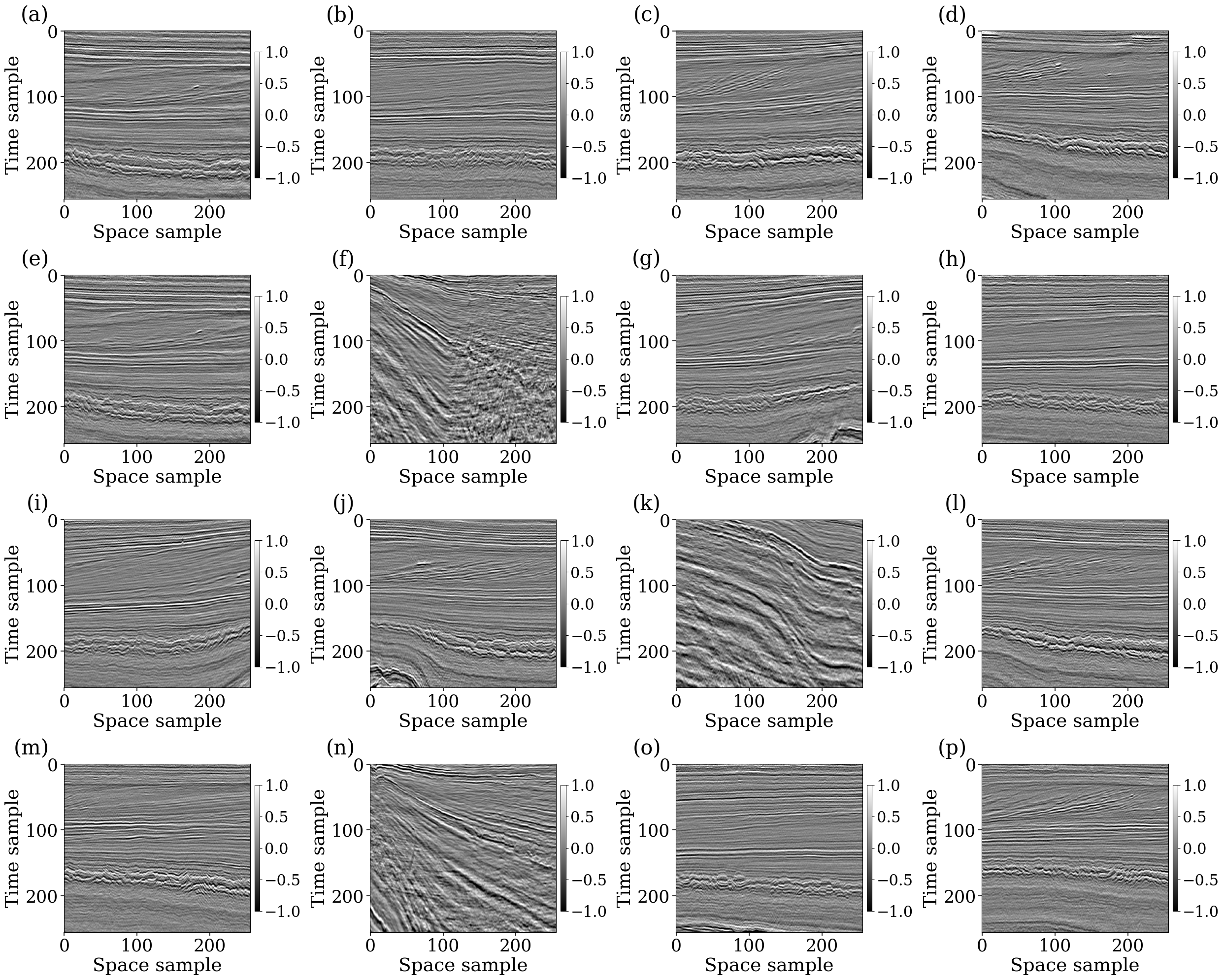}
    \caption{Extraction of the normalized 2D training patches from the field seismic volumes. The volume is sliced along the inline and crossline directions, and \((256 \times 256)\) patches are extracted with a stride of \(128\) samples, yielding 16,165 overlapping patches after filtering patches with excessive zeros. These patches capture a range of varying amplitude characteristics and structural features, including faulted reflectors, continuous horizons, and dipping reflectors.}
    \label{fig:training_patches}
\end{figure}

For conditional training, we employed synthetic seismic data and their corresponding fault label from \cite{wu2019faultseg3d}. However, a key challenge in transferring from field-pretrained to synthetic-conditioned models is the domain gap between simulation outputs and field seismic data. Forward-modeled synthetic seismic data provides accurate fault geometry but exhibits unrealistic textures, including overly smooth reflectors and simplified noise. We address this through NST, following the work of \cite{jun2022realistic}.

\begin{figure}[tbp]
    \centering
    \includegraphics[width=1.0\linewidth]{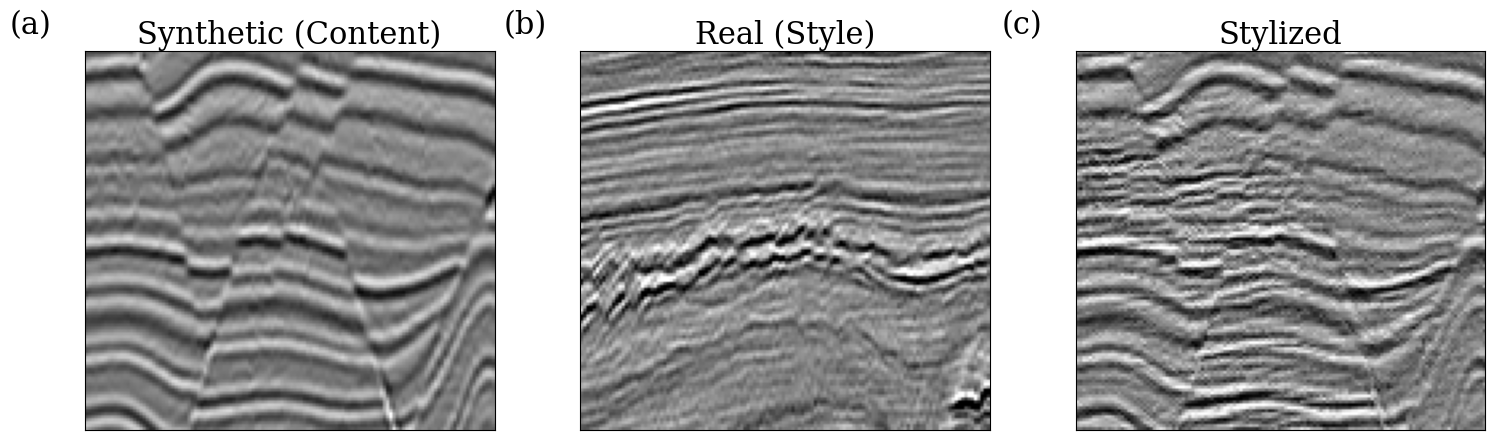}
    \caption{Application of neural style transfer (NST) to bridge the domain gap between synthetic and field seismic data. (a) Content images: synthetic seismic sections from FaultSeg3D with known fault geometry. (b) Style reference: a representative F3 field patch providing the target textural properties. (c) Stylized output: the resulting images after NST, which preserve the structural content (fault locations and geometry) of the synthetic data while adopting the textural and noise characteristics of the field (style) data. The stylized pairs are used as training inputs for conditional fine-tuning of FC-DDPM.}
    \label{fig:NST-enhanced-image}
\end{figure}

Given a field seismic patch, otherwise referred to as the style image, and a synthetic seismic image, referred to as the content image, we optimize for an image (stylized) that preserves synthetic structural content while adopting field textural characteristics. The loss function combines a style loss (matching texture statistics via Gram matrices) and a content loss (preserving fault structure via feature activations). The stylized synthetic dataset retains accurate fault labels from simulation while exhibiting field-realistic textures (see Figure \ref{fig:NST-enhanced-image}). While these textural changes can appear subtle in the image domain, they correspond to a measurable migration of the spectral content toward field statistics, which we quantify in the Discussion (Table~\ref{tab:nst_metrics}).

\section{Model training}
The goal of DDPM is to learn the reverse (denoising) process that gradually denoises \(x_T \approx \mathcal{N}(0,I)\) back to a realistic seismic image \(x_0\). In this work, the reverse process is parameterized using equations~\ref{eq:6} and~\ref{eq:cfg} to predict the injected noise \(\boldsymbol{\epsilon}\) from the noisy input \(\mathbf{x}_t\), the diffusion timestep \(t\), and the optional fault condition \(\mathbf{f}\).

The denoising network used in this work to parameterize both the unconditional and conditional noise predictions is the popular U-NET model, which has been extensively applied to various image segmentation tasks \cite[]{bodapati2023efficient,wang2023enhanced,li2022epick}. The U-NET model is characterized by a downsampling (encoder) path that extracts hierarchical features at progressively coarser spatial resolutions, a bottleneck capturing global contextual information, an upsampling (decoder) path that reconstructs fine-scale structure (i.e., upsample the downsampled features), and symmetric skip connections that help to preserve high-frequency spatial details.

\subsection{UNET architecture}
\label{subsec:UNET}
The network consists of three main components: (1) a time embedding module, (2) a multi-scale encoder–decoder backbone with residual blocks, and (3) a conditional input mechanism that supports classifier-free guidance. The input to the network is a noisy seismic image \( x_t \in \mathbb{R}^{B \times C \times H \times W} \), where \( B \) denotes the batch size and \( C=1 \) is the number of seismic channels. When conditioning is enabled, a fault mask \( f \in \mathbb{R}^{B \times C_f \times H \times W} \) is concatenated channel-wise with the input image, yielding an effective input dimension of \( C + C_f \). The input is augmented with a time embedding \(e_t\) that encodes diffusion time information. 

\subsubsection{Time embedding}
To explicitly encode the diffusion timestep, the scalar timestep \( t \) is embedded using a sinusoidal positional encoding followed by a learnable multilayer perceptron (MLP). The initial sinusoidal embedding maps the timestep to a fixed-dimensional representation using sine and cosine functions with exponentially increasing frequencies, following
\begin{equation}
\gamma(t) = \left[\sin(\omega_k t), \cos(\omega_k t)\right]_{k=1}^{d/2},
\end{equation}
where \( d \) is the embedding dimension. This representation is then processed by a two-layer MLP with SiLU activations to produce the final time embedding \( \tau(t) \in \mathbb{R}^{d} \). The resulting time embedding is injected into each residual block via a learned affine transformation, allowing the network to adapt its denoising behavior continuously across diffusion timesteps.

\subsubsection{Residual blocks}
Residual learning stabilizes training and facilitates gradient flow, which is critical for deep diffusion models operating over hundreds or thousands of timesteps \cite[]{zhang2024residual}. We applied feature extraction at each resolution level using residual blocks. Each block consists of two convolutional layers with \( 3 \times 3 \) kernels, interleaved with Group Normalization and SiLU activations. Given an input feature map \( h \), the residual block computes
\begin{equation}
\mathrm{ResBlock}(h, t) =
h + \mathcal{F}(h, \tau(t)),
\end{equation}
where \( \mathcal{F} \) denotes a nonlinear transformation that includes additive modulation by the time embedding. When the input and output channel dimensions differ, a \( 1 \times 1 \) convolution is applied to the skip connection to ensure dimensional consistency.

\subsubsection{Encoder (downsampling path)}
The encoder progressively reduces the spatial resolution while increasing the number of feature channels, enabling the network to capture increasingly abstract representations. Downsampling is implemented using strided \( 4 \times 4 \) convolutions with stride 2. At each resolution, a residual block processes the features before downsampling. This hierarchical representation allows the model to encode long-range spatial dependencies and large-scale geological structures, such as fault trends and reflector continuity.

\subsubsection{Bottleneck}

At the lowest resolution, two residual blocks are applied without further downsampling. This bottleneck stage operates at the highest feature dimensionality and serves as the primary integration point for global contextual information. Time embeddings are injected at this stage in the same manner as in the encoder and decoder blocks.

\subsubsection{Decoder (upsampling path)}
The decoder mirrors the encoder structure, progressively increasing spatial resolution through transposed convolutions. At each resolution level, the upsampled features are concatenated with the corresponding encoder features via skip connections. This concatenation preserves high-resolution spatial details that may be lost during downsampling. Following concatenation, residual blocks fuse the multi-scale information, allowing the network to jointly reason about fine-scale seismic textures and large-scale fault geometry.

\subsubsection{Conditional input and classifier-free guidance}
Conditioning on fault geometry is implemented by concatenating the fault mask with the noisy seismic input at the network entrance. During training, classifier-free guidance is enabled by randomly dropping the conditioning input with a fixed probability and replacing it with zeros. This allows the same network to learn both conditional and unconditional noise estimators. At inference time, the conditional and unconditional predictions can be linearly combined to control the conditioning strength via a guidance scale parameter, thereby enabling a trade-off between strict fault conformity and preservation of realistic seismic texture.

\begin{figure}[tbp]
    \centering
    \includegraphics[width=1.0\linewidth]{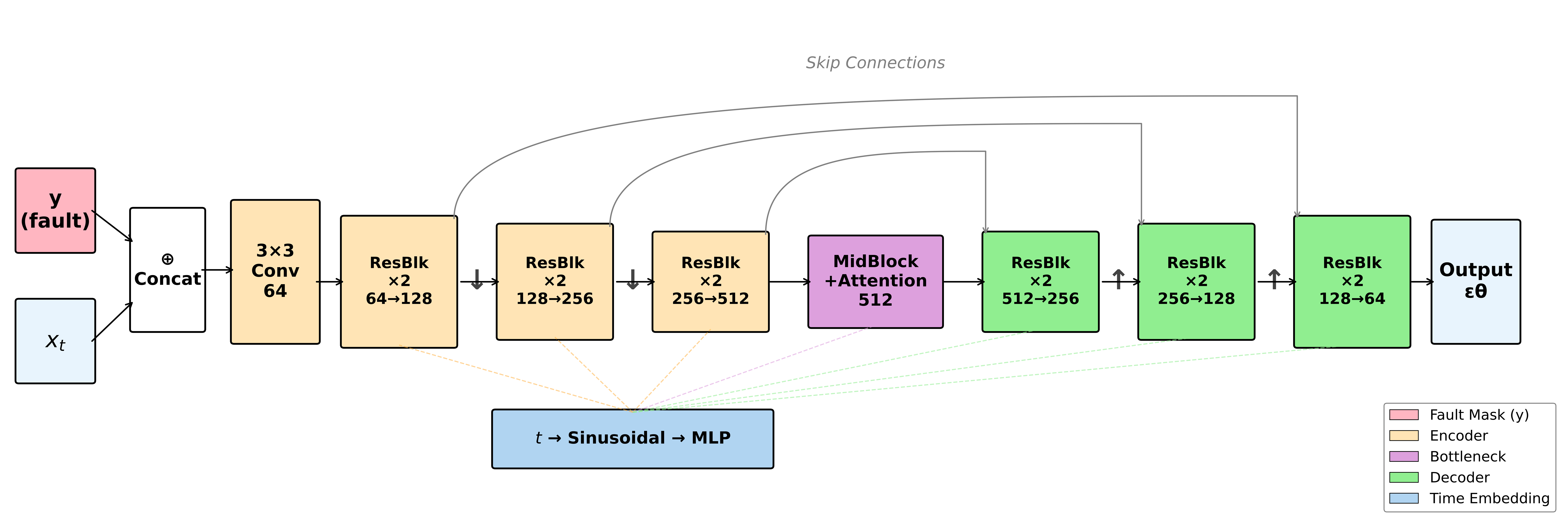}
    \caption{U-Net architecture for noise prediction in U-DDPM and FC-DDPM. The encoder compresses the spatial input along the channel dimension with strides of 64, 128, 256, and 512 using strided convolutions. Each resolution contains two residual blocks \((\text{ResBlk} \times 2)\) with Group Normalization and SiLU activation. Self-attention is applied at the 512-channel bottleneck. The decoder mirrors the encoder with transposed convolutions and concatenative skip connections at each resolution. Sinusoidal timestep embeddings are projected through a multi-layer perceptron (MLP) and injected into every residual block. For conditional generation, the binary fault mask \(f\) is concatenated with the noisy input \(x_t\) along the channel dimension, yielding a two-channel input to the initial \(3 \times 3\) convolution.}
    \label{fig:unet}
\end{figure}

\subsubsection{Output layer}
The final output layer consists of a Group Normalization layer followed by a \( 3 \times 3 \) convolution that maps the decoder features to a single-channel output. The network predicts the unconditional and conditional noise component, i.e., \(\epsilon_\theta(x_t, t)\) and \( \epsilon_\theta(x_t, t, f) \), which are used to parameterize the reverse diffusion transition for both unconditional and conditional training processes. An overview of the architecture is shown conceptually in Figure~\ref{fig:unet}. A summary of the training hyperparameters is presented in Table \ref{tab:hyperparameters}.

\begin{table}[t]
\centering
\caption{Training hyperparameters and computational costs for U-DDPM and FC-DDPM}
\label{tab:hyperparameters}
\begin{tabular}{lcc}
\hline
\textbf{Parameter} & \textbf{U-DDPM} & \textbf{FC-DDPM} \\
\hline
Total diffusion steps $T$ & 1000 & 1000 \\
Training iterations & 50{,}000 & 15{,}000 \\
Batch size & 16 & 16 \\
Learning rate & $2 \times 10^{-4}$ & $5 \times 10^{-5}$ \\
Optimizer & AdamW & AdamW \\
EMA decay & 0.999 & 0.999 \\
Noise schedule & Cosine & Cosine \\
CFG dropout $p_{\mathrm{uncond}}$ & -- & 0.2 \\
Guidance scale, $s$ (sampling) & -- & 3.0 \\
\hline
Hardware & \multicolumn{2}{c}{NVIDIA A100-80GB (single GPU)} \\
Training time (approx.) & $\sim$ 2 h & $\sim$0.4 h \\
Inference (per patch) & $\sim$5.8 s ($256^2$) & $\sim$2.7 s ($128^2$) \\
\hline
\end{tabular}
\end{table}

\paragraph{Computational efficiency.} All experiments were conducted on a single NVIDIA A100-80GB GPU. Unconditional pretraining (50{,}000 iterations at $256\times256$) required approximately 2~h, and conditional fine-tuning (15{,}000 iterations at $128\times128$) approximately 0.4~h, for a total training cost under three GPU-hours. At inference, FC-DDPM generates a $128\times128$ fault-conditioned patch in $\sim$2.7~s using the full 1{,}000-step reverse process; throughput scales with batch size, and accelerated samplers (e.g., DDIM) can further reduce this cost. Crucially, generation is a one-time, fully amortized expense: once trained, an arbitrary number of realizations can be sampled from any fault mask without further optimization or survey-specific parameter tuning, in contrast to physics-based forward modeling.


\section{seismic image generation}
\subsection{Unconditional sampling (U-DDPM)}
\begin{figure}[tbp]
    \centering
    \includegraphics[width=1\linewidth]{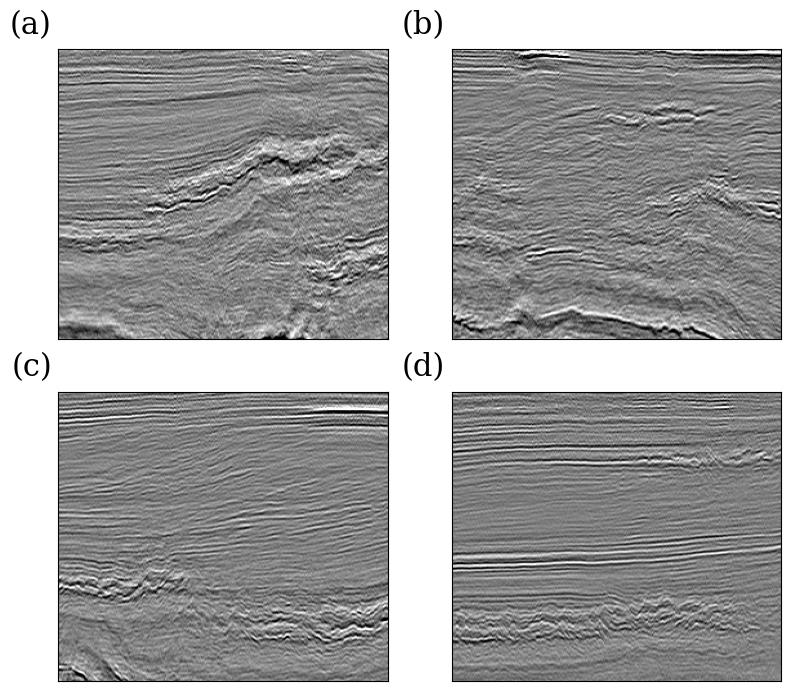}
    \caption{Seismic images generated unconditionally by U-DDPM after pretraining on F3 field patches. Each \(256 \times 256\) sample is produced by running the reverse diffusion process over 1,000 timesteps from Gaussian noise using EMA-averaged weights \((\gamma = 0.999)\) . The generated images display coherent reflector continuity, amplitude variations, and structural diversity consistent with the field data.}
    \label{fig:U-DDPM samples}
\end{figure}

After training, new seismic images are generated by reverse-diffusion sampling using the fixed EMA parameters \(\boldsymbol{\theta}_{\mathrm{EMA}}\) defined in equation~\ref{eq:ema}. Starting from \(\mathbf{x}_T\sim\mathcal{N}(\mathbf{0},\mathbf{I})\), we apply the reverse update in equation~\ref{eq:6} for \(t=T,\ldots,1\), with \(T=1000\), using the EMA parameters in place of the training parameters. The procedure is summarized in Algorithm~\ref{alg:uddpm_sampling}.
Examples of generated seismic images without any imposed condition are shown in Figure~\ref{fig:U-DDPM samples}.

\begin{figure}[tbp]
    \centering
    \includegraphics[width=1.0\linewidth]{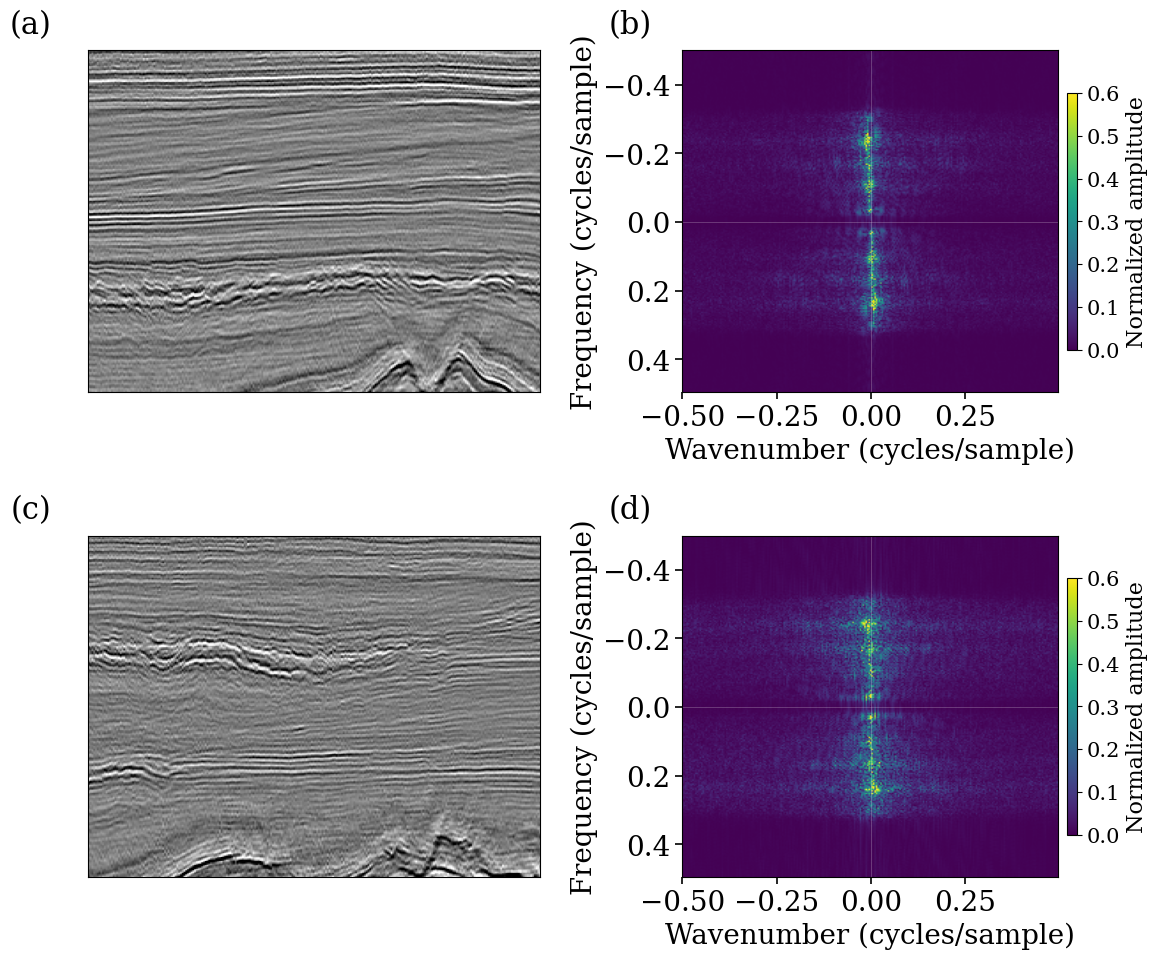}
    \caption{Evaluation of generated images using unconditional denoising diffusion probabilistic model (U-DDPM). (a) A patch of field seismic data used for training. (b) Frequency-wavenumber (F-K) amplitude spectrum of the field seismic data used. (c) Seismic image generated via U-DDPM. (d) F-K amplitude spectrum of unconditionally generated data. The generated data reproduces the spectral bandwidth, dip distribution, and energy concentration of the field data, providing quantitative confirmation that the model has learned the spatio-temporal statistics of the training set beyond visual resemblance.}
    \label{fig:F-K-GEN-U-DDPM}
\end{figure}

To move beyond visual inspection and evaluate whether the U-DDPM reproduces the physical and textural characteristics of the training real seismic data, we analyze both the original and generated seismic sections in the frequency (F-K) domain. F-K analysis provides a joint representation of seismic energy as a function of temporal frequency and spatial wavenumber, enabling quantitative assessment of spectral bandwidth, dip content, and spatio-temporal correlations that are not easily diagnosed in the time (t-x) domain alone. Given a seismic section \(s(t,x)\), where \(t\) represents the time component and \(x\) represents the distance component, the F-K spectrum is obtained via a two-dimensional Fourier transform,
\begin{equation}
    S(f,k) = \mathcal{F}_{2D}{s(t,x)}
\end{equation}
where \(f\) is the temporal frequency and \(k\) is spatial wavenumber. The magnitude \(|S(f,k)|\) represents the energy content of the seismic data at each frequency-wavenumber pair. Comparing the F-K spectra for the field and generated seismic realizations helps to assess whether the model has learned the joint spatio-temporal statistics of the training data, rather than merely reproducing visually plausible textures. Figure \ref{fig:F-K-GEN-U-DDPM} illustrate an agreement between the generated and field seismic data, indicating the model preserves key physical characteristics, including temporal bandwidth, spatial correlation, and dip distribution. 

\begin{algorithm}[t]
\caption{U-DDPM Sampling (Unconditional Generation)}
\label{alg:uddpm_sampling}
\begin{algorithmic}[1]
\REQUIRE Trained model $\boldsymbol{\epsilon}_\theta$ with fixed EMA parameters, 
schedule $\{\beta_t, \alpha_t, \bar{\alpha}_t\}$
\ENSURE Generated seismic image $\mathbf{x}_0$
\STATE Sample $\mathbf{x}_T \sim \mathcal{N}(\mathbf{0}, \mathbf{I})$
\FOR{$t = T, T-1, \ldots, 1$}
    \STATE Sample $\mathbf{z} \sim \mathcal{N}(\mathbf{0}, \mathbf{I})$ if $t > 1$, else $\mathbf{z} = \mathbf{0}$
    \STATE $\hat{\boldsymbol{\epsilon}} \leftarrow \boldsymbol{\epsilon}_\theta(\mathbf{x}_t, t)$
    \STATE $\boldsymbol{\mu}_t \leftarrow \frac{1}{\sqrt{\alpha_t}} \left( \mathbf{x}_t - \frac{\beta_t}{\sqrt{1 - \bar{\alpha}_t}} \hat{\boldsymbol{\epsilon}} \right)$
    \STATE $\mathbf{x}_{t-1} \leftarrow \boldsymbol{\mu}_t + \sqrt{\beta_t} \cdot \mathbf{z}$
\ENDFOR
\RETURN $\mathbf{x}_0$
\end{algorithmic}
\end{algorithm}

\subsection {Imposing a fault condition (FC-DDPM)}
\label{sec:fault_condition}
During the sampling phase, we impose a condition on the image-generation process. Specifically, we want the generated seismic images to honor a predefined fault geometry. We imposed this condition for two main reasons: (1) to enforce structural consistency and geological plausibility in our generated seismic images. U-DDPM learns the marginal distribution of seismic images but lacks an explicit mechanism to preserve specific structural features, such as fault location, orientation, or reflector continuity. By imposing a fault condition, the generative process is constrained to honor a prescribed fault geometry. (2) to enhance data augmentation for DL-based fault detection (DL-FD) tasks. DL-FD tasks are highly sensitive to the diversity and quantity of fault examples available in the training data. Imposing a fault condition during DDPM sampling enables targeted generation of seismic images with diverse fault attributes, thereby improving the balance and coverage of fault-related features in the training data and leading to more robust, generalizable fault-detection models.

\begin{figure}[tbp]
    \centering
    \includegraphics[width=1\linewidth]{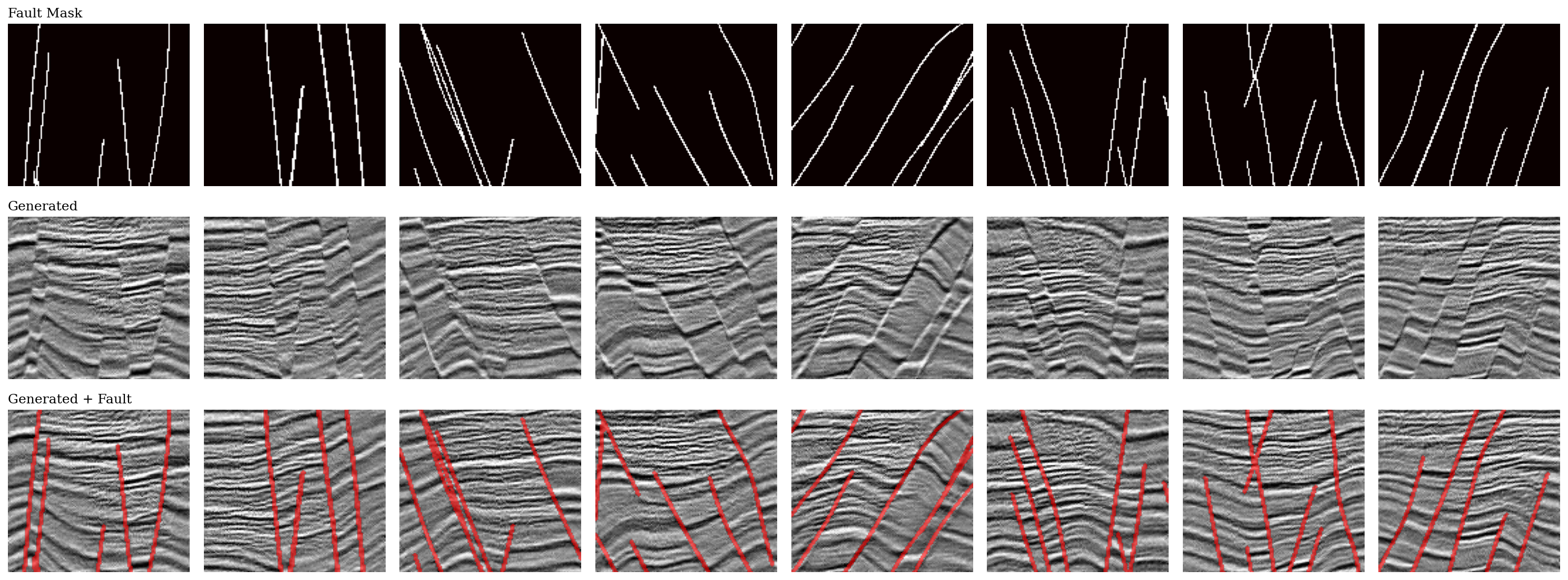}
    \caption{Fault-conditioned seismic images generated by FC-DDPM, demonstrating adherence to the imposed fault geometry. \textbf{Top row:} eight binary fault masks \(\mathbf{f}\) used as conditioning inputs, spanning a range of fault densities, dips, and intersection patterns. \textbf{Middle row:} the corresponding seismic images generated by FC-DDPM using classifier-free guidance \((s = 3.0)\); each is an independent realization sampled from Gaussian noise and conditioned \emph{only} on the mask above it, not reconstructed from any reference image. \textbf{Bottom row:} the same generated images with the input fault mask overlaid in red. The generated reflector terminations and amplitude discontinuities coincide with the imposed fault traces, confirming that the model honors the prescribed fault locations, orientations, and dips while synthesizing field-like reflector textures around them. Quantitative generation-quality metrics are reported in Table~\ref{tab:gen_quality}.}
    \label{fig:fc-ddpm gens}
\end{figure}

The conditioning process is incorporated exclusively during the reverse "denoising" stage. Given a fault mask, \(f\), the Gaussian translation that needs to be learned at every timestep, i.e., \(p_\theta(x_0 | x_T, f)\), is conducted using equation \ref{eq:9}. FC-DDPM is structured so that the model generates a target image \(x_0\), starting with a completely noisy image, \(x_T\), and iteratively refines the output image to obtain a sequence \((x_{T-1}, x_{T-2}, ..., X_0)\) according to the learned conditional distributions \(p_\theta(x_0|x_T, f)\), such that \(x_0 \approx p(x|f) \) \cite[]{saharia2022image}. The trained unconditional DDPM on field data had enabled the model to learn realistic amplitude distributions, frequency content, and spatial correlations inherent to real seismic images. Subsequently, the learned weights of the U-Net backbone are transferred to initialize a fault-conditioned DDPM trained on stylized synthetic seismic images and corresponding fault labels. Figure \ref{fig:fc-ddpm gens} shows the generated seismic images given a prescribed fault geometry. 

\begin{figure}[tbp]
    \centering
    \includegraphics[width=1\linewidth]{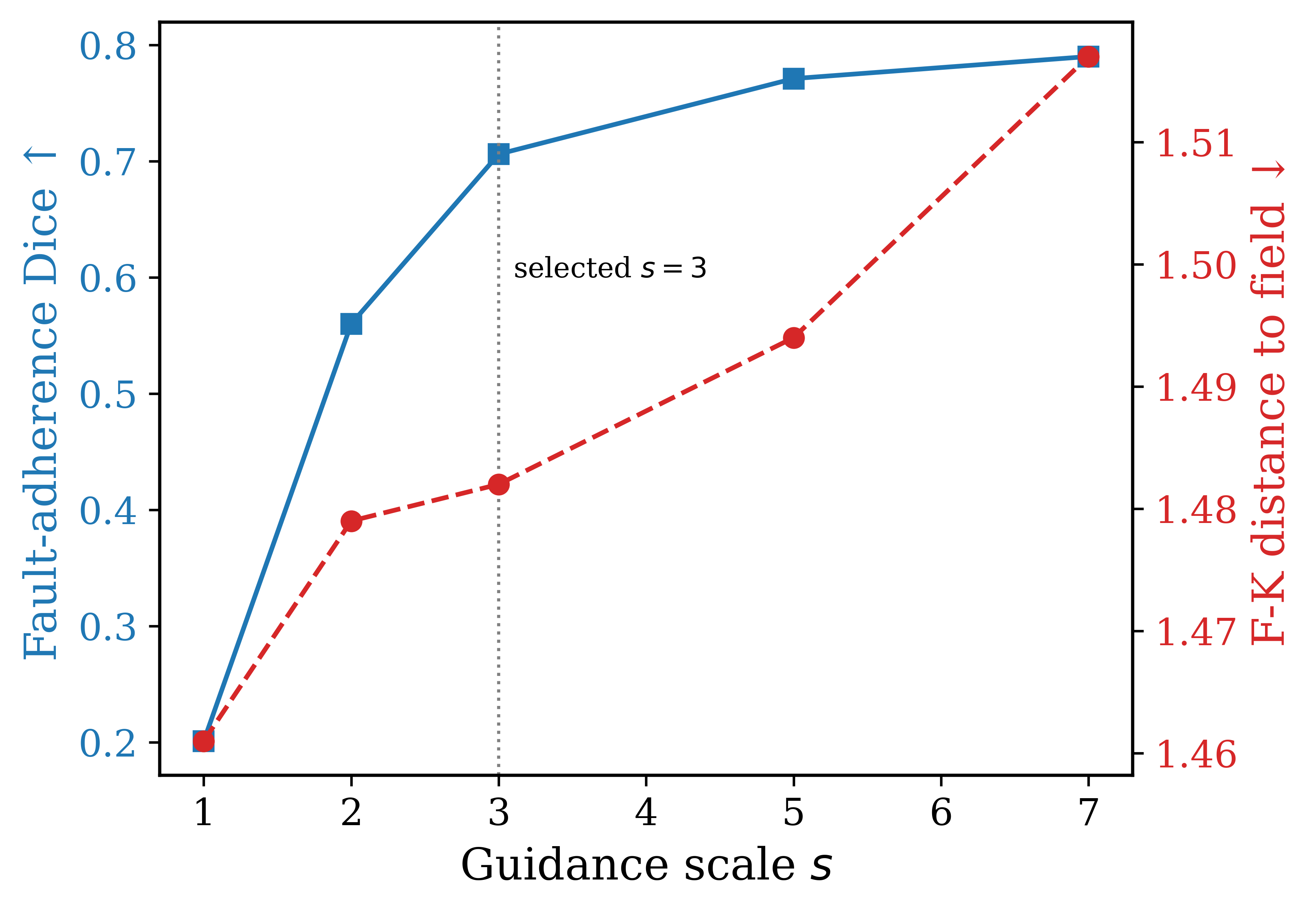}
    \caption{Sensitivity of FC-DDPM conditional generation to the classifier-free guidance scale $s$. Fault adherence (blue, left axis) is the Dice coefficient between an independent, synthetic-trained fault detector's prediction on the generated image and the input fault mask; it increases sharply from $s=1$ (Dice $0.20$) to $s=3$ (Dice $0.71$) and saturates thereafter ($0.79$ at $s=7$). Textural realism, measured by the frequency--wavenumber (F-K) spectral distance to field data (red, right axis; lower is more field-like), degrades modestly and monotonically with increasing $s$ ($1.46 \rightarrow 1.52$), as guidance progressively extrapolates the noise estimate beyond the conditional prediction and off the learned data manifold. The selected value $s=3$ (dotted line) lies at the knee of this trade-off, achieving approximately $90\%$ of the maximum attainable fault fidelity while preserving textural realism and sample diversity.}
    \label{fig:guidance}
\end{figure}

\paragraph{Guidance scale.} The classifier-free guidance scale ($s$) controls the trade-off between fault-conditioning fidelity and textural realism. Figure~\ref{fig:guidance} reports a sensitivity analysis over $s \in \{1,2,3,5,7\}$, quantifying fault adherence (Dice between an independent detector's prediction on the generation and the input mask) and realism (F-K distance to field). At low guidance ($s=1$), the model largely ignores the mask (Dice $0.20$) while generating realistic texture; adherence then rises sharply to $s=3$ (Dice $0.71$) and saturates thereafter ($0.79$ at $s=7$). Realism degrades modestly and monotonically as $s$ increases (F-K distance $1.46 \rightarrow 1.52$), because guidance extrapolates the noise estimate beyond the conditional prediction and progressively pushes samples off the learned data manifold, which also reduces sample diversity. We therefore adopt $s=3$, at the knee of this trade-off: it captures approximately 90\% of the maximum attainable adherence while retaining near-optimal realism and diversity. This value was used for all reported generations and downstream experiments.

We assess the quality of the FC-DDPM generations against field data along three axes (Table~\ref{tab:gen_quality}). \emph{Spectrally}, the frequency--wavenumber (F-K) distance of the generations to the field distribution ($1.48$) is substantially smaller than that of the raw synthetic data ($1.79$), indicating that the generations acquire field-like dip content and spatial correlations. \emph{Statistically}, the generated amplitude distribution closely matches field (Wasserstein distance $0.07$), with slightly lighter tails (excess kurtosis $-0.2$ vs.\ $0.6$), indicating that the domain adaptation acts on spectral and spatial texture rather than the amplitude histogram. \emph{Structurally}, an independent, synthetic-trained fault detector recovers the imposed fault geometry from the generations with a Dice of $0.70$ --- exceeding the fault detectability of the stylized training data ($0.34$) --- confirming that FC-DDPM produces clear, well-localized fault expressions consistent with the conditioning masks.

\begin{table}[tbp]
\centering
\caption{Quantitative evaluation of FC-DDPM generations against field seismic data across spectral, statistical, and structural axes. Amplitude and F-K metrics use consistently scaled patches; fault-adherence Dice uses an independent synthetic-trained detector.}
\label{tab:gen_quality}
\begin{tabularx}{\linewidth}{@{}lYr@{}}
\hline
\textbf{Axis} & \textbf{Metric} & \textbf{Value} \\
\hline
Spectral    & F-K distance to field (synthetic $\to$ generated) & $1.79 \to 1.48$ \\
Statistical & Amplitude Wasserstein distance to field           & $0.07$ \\
Statistical & Excess kurtosis (field $=0.56$)                   & $-0.21$ \\
Structural  & Fault-adherence Dice (training-data ref.\ $=0.34$) & $0.70$ \\
\hline
\end{tabularx}
\end{table}

\begin{algorithm}[t]
\caption{FC-DDPM Sampling with Classifier-Free Guidance}
\label{alg:fcddpm_sampling}
\begin{algorithmic}[1]
\REQUIRE Trained model $\boldsymbol{\epsilon}_\theta$, 
with fixed EMA parameters,
fault mask $\mathbf{f}$, guidance scale $s$
\REQUIRE Schedule $\{\beta_t, \alpha_t, \bar{\alpha}_t\}$
\ENSURE Fault-conditioned seismic image $\mathbf{x}_0$
\STATE Sample $\mathbf{x}_T \sim \mathcal{N}(\mathbf{0}, \mathbf{I})$
\FOR{$t = T, T-1, \ldots, 1$}
    \STATE Sample $\mathbf{z} \sim \mathcal{N}(\mathbf{0}, \mathbf{I})$ if $t > 1$, else $\mathbf{z} = \mathbf{0}$
    \STATE $\hat{\boldsymbol{\epsilon}}_{\text{cond}}
    \leftarrow \boldsymbol{\epsilon}_\theta^{\mathrm{CFG}}(\mathbf{x}_t, t, \mathbf{f})$
    \STATE $\hat{\boldsymbol{\epsilon}}_{\text{uncond}}
    \leftarrow \boldsymbol{\epsilon}_\theta(\mathbf{x}_t, t, \mathbf{0})$
    \STATE $\hat{\boldsymbol{\epsilon}} \leftarrow \hat{\boldsymbol{\epsilon}}_{\text{uncond}} + s \cdot \left( \hat{\boldsymbol{\epsilon}}_{\text{cond}} - \hat{\boldsymbol{\epsilon}}_{\text{uncond}} \right)$
    \STATE $\boldsymbol{\mu}_t \leftarrow \frac{1}{\sqrt{\alpha_t}} \left( \mathbf{x}_t - \frac{\beta_t}{\sqrt{1 - \bar{\alpha}_t}} \hat{\boldsymbol{\epsilon}} \right)$
    \STATE $\mathbf{x}_{t-1} \leftarrow \boldsymbol{\mu}_t + \sqrt{\beta_t} \cdot \mathbf{z}$
\ENDFOR
\RETURN $\mathbf{x}_0$
\end{algorithmic}
\end{algorithm}

\section{Discussion}
The experimental results demonstrate that the unconditional and fault-conditioned denoising diffusion probabilistic model (U-DDPM and FC-DDPM) successfully generate seismic images that exhibit both realistic field-like textures and strict adherence to prescribed fault geometries. This dual capability addresses a fundamental challenge in DL-FD tasks. The F-K spectral analysis, shown in Figure \ref{fig:F-K-GEN-U-DDPM}, provides quantitative evidence that generated images preserve the spatio-temporal statistics of field seismic data, including temporal bandwidth and spatial correlation. 

The transfer learning strategy employed in this work proves essential for achieving efficient training. Pretraining on unlabeled data, followed by conditional fine-tuning on stylized synthetic data, enables the network to first learn realistic seismic textures before fine-tuning for structural conditioning. In contrast to the unconditional training run for 50,000 timesteps, the conditional training began producing realistic seismic images at 1000 steps, indicating that the network leverages learned seismic priors rather than learning texture representations from scratch. 

\begin{figure}[tbp]
    \centering
    \includegraphics[width=1\linewidth]{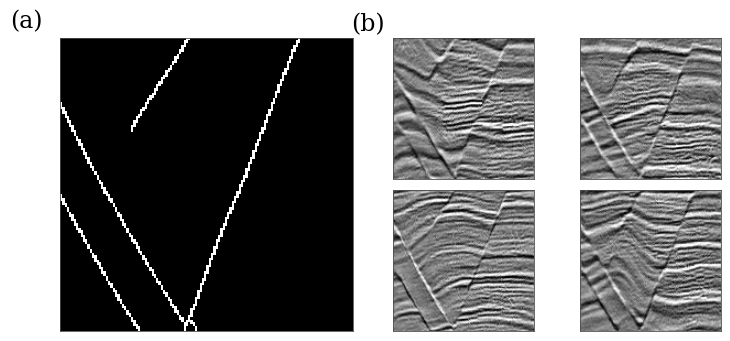}
    \caption{One-to-many conditional sampling from FC-DDPM. (a): a single input fault mask. (b) seismic realizations generated from the same fault condition using different random seeds. Each sample exhibits the same fault geometry but with distinct reflector patterns and amplitude characteristics, demonstrating the stochastic diversity of the diffusion sampling process and the model's suitability for generating varied training data from limited structural labels.}
    \label{fig:OneFault-ManySeismic}
\end{figure}

We also demonstrate the capability of one-to-many conditional sampling: given a single fault geometry, we can generate an unlimited number of seismic data with varying textures that exhibit the same imposed fault geometry, as illustrated in Figure \ref{fig:OneFault-ManySeismic}. This enables the generation of diverse training data for DL-based fault detection tasks.

\subsection{Role of Style Transfer in Domain Bridging}
\label{subsec:nst_role}
A key challenge in conditioning seismic generation on faults is the lack of paired fault labels in field seismic data. To address this limitation, we adopt conditional fine-tuning on synthetic seismic data with fault-pair labels. Visual inspection of the initial fine-tuning reveals that the generated images exhibit a textural resemblance similar to that of the synthetic seismic data. This is because the model cannot learn fault structure from synthetic data without simultaneously learning synthetic texture from seismic data. To address this, we applied a style transfer approach that transferred the textural characteristics of the F3 data to the 2000 synthetic seismic patches. Integrating NST helps us to bridge the gap between field and synthetic data. 

The F-K spectral comparison shown in Figure \ref{fig:NST_F-K_Spectrum} reveals the effect of this domain bridging. The stylized data inherit the vertically elongated amplitude spectrum characteristics of the field data, indicating broader dip diversity and more realistic spatial correlations. Consequently, FC-DDPM trained on stylized data generates images whose spectral properties closely match those of field data, validating NST as an effective domain adaptation technique for seismic applications and further extending the work of \cite{jun2022realistic} by showing that style-transferred data can serve as effective conditioning inputs for generative models. 

\paragraph{Quantitative assessment of style transfer.} To move beyond visual inspection, we quantify the effect of NST on both structure and texture over the $N = 2000$ synthetic--stylized training pairs (Table~\ref{tab:nst_metrics}). Decomposing each image into low- and high-frequency components, NST preserves the large-scale structural layout, i.e., the reflector configuration and fault geometry, with a low-pass structural similarity index (SSIM) of $0.82 \pm 0.08$, while more substantially modifying the fine reflector texture (high-pass SSIM of $0.61 \pm 0.06$). This frequency-selective behavior is the desired outcome: the style transfer adapts texture while retaining the structural content on which fault conditioning depends. To verify that the adaptation is meaningful rather than cosmetic, we compute the frequency--wavenumber (F-K) spectral distance of the synthetic and stylized patches to the field distribution. NST reduces the mean F-K distance to field by $21.7\%$ (from $1.91$ to $1.49$), quantitatively confirming the spectral migration toward field statistics that Figure~\ref{fig:NST_F-K_Spectrum} shows qualitatively. Although the texture change can appear subtle in the image domain, this result demonstrates that the stylized data acquire field-like spatio-temporal statistics while preserving geological structure.

\begin{table}[t]
\centering
\caption{Quantitative assessment of neural style transfer (NST) over $N = 2000$ synthetic--stylized pairs. Low- and high-pass SSIM separate large-scale structure from fine texture; the frequency--wavenumber (F-K) distance measures spectral proximity to the field distribution (lower is closer).}
\label{tab:nst_metrics}
\begin{tabularx}{\linewidth}{@{}YcY@{}}
\hline
\textbf{Metric} & \textbf{Value} & \textbf{Interpretation} \\
\hline
Low-pass SSIM (structure)        & $0.82 \pm 0.08$ & Large-scale geometry preserved \\
High-pass SSIM (texture)         & $0.61 \pm 0.06$ & Fine texture selectively modified \\
F-K distance to field, synthetic & $1.79$          & Baseline spectral gap \\
F-K distance to field, stylized  & $1.48$          & $21.7\%$ closer to field \\
\hline
\end{tabularx}
\end{table}

\begin{figure}[tbp]
    \centering
    \includegraphics[width=1\linewidth]{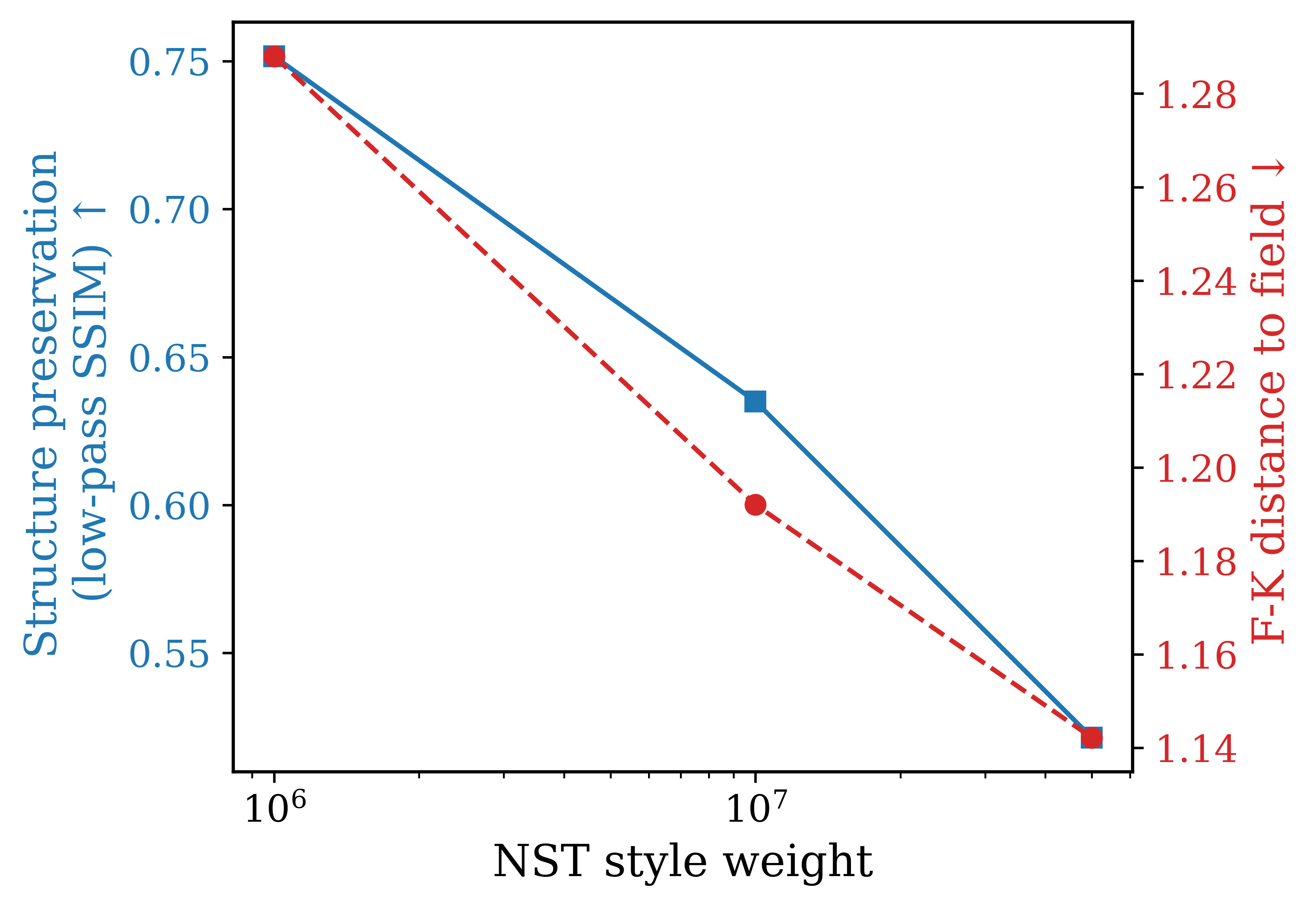}
    \caption{Effect of the neural style-transfer weight on the stylized training data. As the style weight increases (log scale), textural realism improves (F-K spectral distance to field --- red, right axis --- decreases), while structure preservation degrades (low-pass structural similarity between synthetic content and stylized output --- blue, left axis --- decreases). A moderate style weight balances textural realism against preservation of reflector and fault geometry.}
    \label{fig:styleweight}
\end{figure}

\begin{figure}[tbp]
    \centering
    \includegraphics[width=1.0\linewidth]{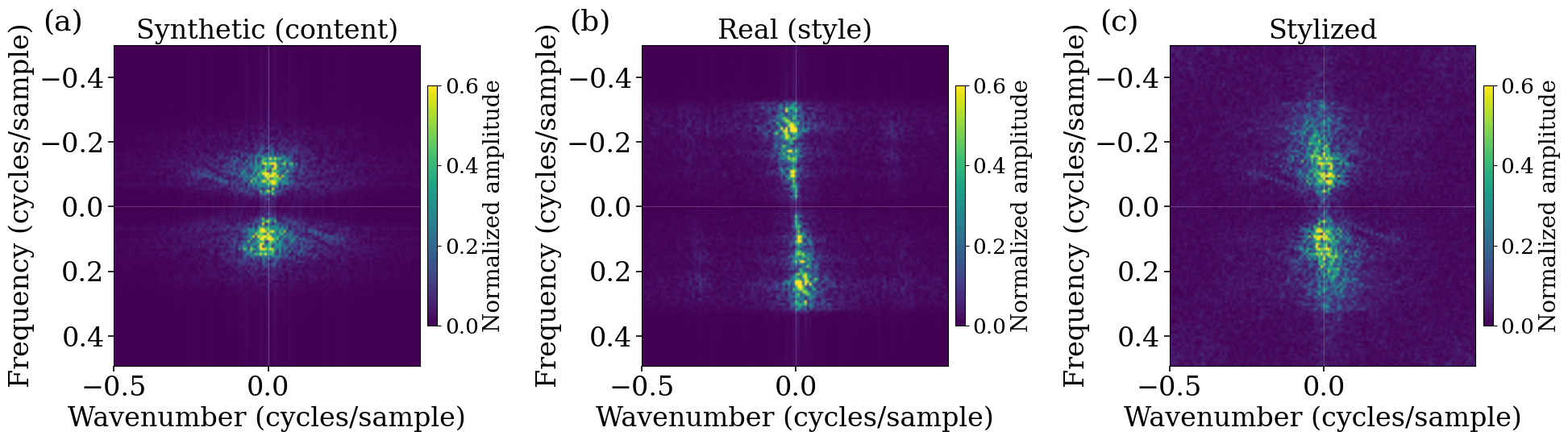}
    \caption{Frequency-wavenumber (F-K) amplitude spectra comparing (a) synthetic seismic data from convolution modeling, (b) F3 field data, and (c) stylized synthetic data after NST. The synthetic data exhibits a compact spectral footprint with limited dip content. The field data shows a broader, vertically elongated spectrum reflecting richer spatial correlations. After style transfer, the stylized data acquires the elongated spectral shape of the field data while maintaining the structural features of the synthetic input, confirming effective transfer of physical characteristics.}
    \label{fig:NST_F-K_Spectrum}
\end{figure}

The NST style weight governs the balance between preserving synthetic structural content and matching field texture. Figure~\ref{fig:styleweight} reports a sensitivity analysis over low, medium, and high style weights. As the style weight increases, the stylized data become more field-like (F-K distance to field decreasing from $1.29$ to $1.15$) but structure preservation degrades (low-pass SSIM decreasing from $0.75$ to $0.52$), as increasingly aggressive style matching begins to distort reflector and fault geometry. We therefore adopt a moderate style weight ($10^{7}$), which captures most of the achievable realism while preserving fault geometry.

\begin{figure}[tbp]
    \centering
    \includegraphics[width=\textwidth]{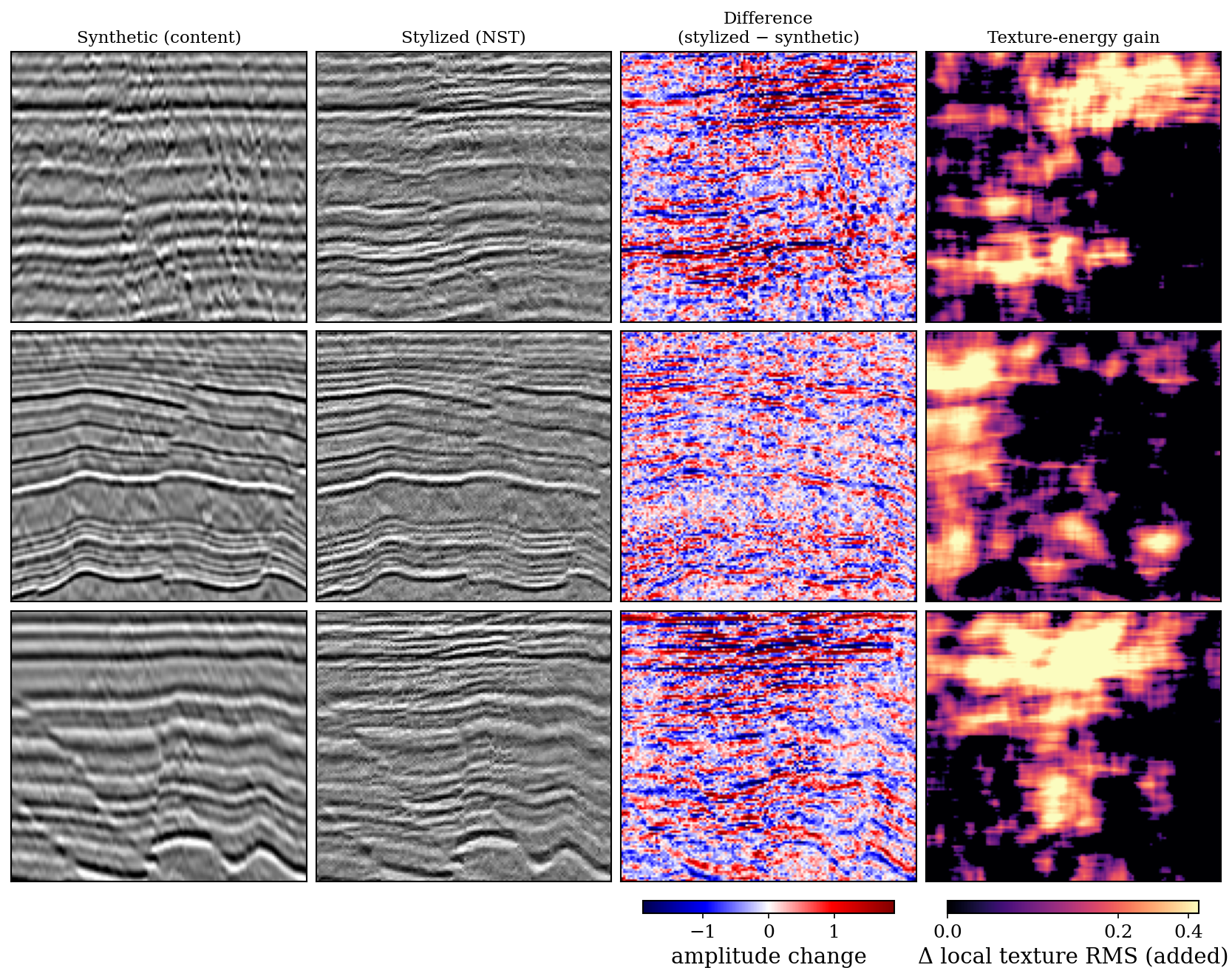}
    \caption{Visualization of the change introduced by neural style transfer (NST). Columns, left to right: synthetic content patch; stylized output; their signed difference (amplitude change); and the texture-energy gain, defined as the increase in local high-frequency RMS amplitude. NST preserves the large-scale reflector and fault geometry while injecting broadband, field-like fine texture (columns 3--4). The difference is spatially incoherent, consistent with added texture and noise rather than coherent acquisition footprint. Texture gains are largest where the synthetic content was smoothest, consistent with the feature-extractor limitations discussed in the text and with the SSIM decomposition (Table~\ref{tab:nst_metrics}) and F-K spectral shift (Figure~\ref{fig:NST_F-K_Spectrum}).}
    \label{fig:nst_difference}
\end{figure}

To further visualize the effect of NST, Figure~\ref{fig:nst_difference} compares representative synthetic content patches with their stylized counterparts. The large-scale reflector geometry and fault structure are preserved (columns 1–2), while the difference (column 3) reveals a substantial, broadband, fine-scale change concentrated on the reflectors rather than a displacement of them. This change is the field-like texture and amplitude modulation injected by NST; it is spatially incoherent and therefore distinct from coherent acquisition-related artifacts such as footprint. The texture-energy gain (column 4), defined as the increase in local high-frequency RMS amplitude, shows that NST enriches fine-scale texture across the section, with the largest gains where the synthetic content was smoothest.

A notable limitation of the NST implementation is its reliance on VGG-19 pretrained on natural images (ImageNet) for feature extraction. The feature representations learned from photographs do not encode geophysically meaningful attributes that govern reflector continuity. Consequently, the Gram matrix-based style matching operates on a feature space that is suboptimal for seismic data, contributing to localized texture inconsistencies in the stylized image, including abrupt amplitude-envelope variations, which then propagate into the FC-DDPM outputs. This imposes an upper bound on the textural fidelity of the generated seismic images, governed by the quality of the style transfer stage. The localized concentration of the texture-energy enrichment is consistent with the limitations of the ImageNet-pretrained feature extractor and complements the quantitative structural/textural SSIM decomposition (Table~\ref{tab:nst_metrics}) and the F-K spectral shift toward field statistics (Figure~\ref{fig:NST_F-K_Spectrum}).Future work could explore training a domain-specific feature extractor on seismic data to better capture geophysically relevant style representations, potentially improving both the local consistency of the stylized training pairs and the downstream generative quality.

\subsection{Downstream Task Validation}
\begin{figure}[tbp]
    \centering
    \includegraphics[width=1.0\linewidth]{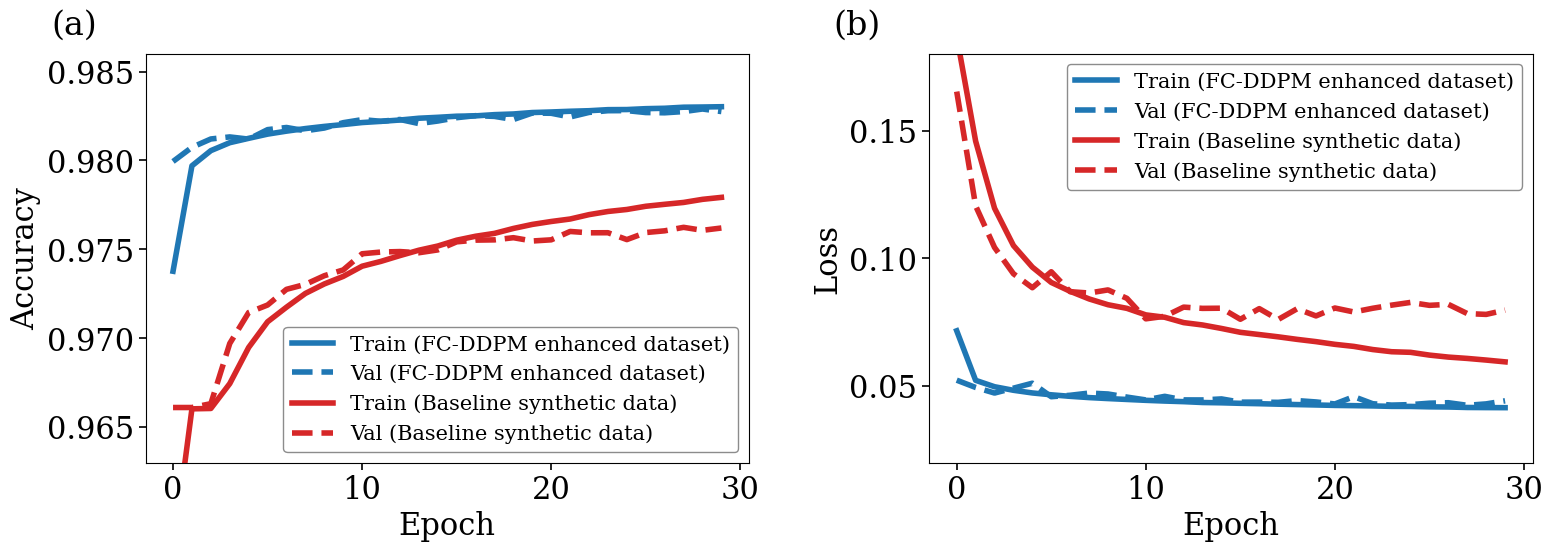}
    \caption{Comparison of the effect of training sample size on the loss function curve for two separate training experiments. (a) Training and validation accuracy curves for a U-Net fault detection model trained on two datasets: 2,000 synthetic samples only (baseline) and the same samples augmented with 20,000 FC-DDPM-generated images. (b) Training and validation loss curves for the two experiments. The augmented training yields lower validation loss and a reduced generalization gap, indicating improved model robustness without any architectural changes.}
    \label{fig:loss_curve_FaultSeg3D_FC-DDPMGen}
\end{figure}

The ultimate test of any data augmentation strategy lies in its impact on downstream task performance. Our experiments with synthetic data demonstrate that augmenting the original 2,000 synthetic training samples with 20,000 FC-DDPM-generated images yields measurable improvements in fault-detection accuracy. More importantly, the augmented training significantly reduces overfitting, as evidenced by the convergence between training and validation loss curves (Figure \ref{fig:loss_curve_FaultSeg3D_FC-DDPMGen}), in contrast to the divergence between the training and validation using only 2000 synthetic dataset. This regularization effect occurs without any modification to the network architecture or training hyperparameters, simply by increasing the training sample size, indicating that the performance gains derive purely from increased data diversity.

\begin{figure}[tbp]
    \centering
    \includegraphics[width=1.0\linewidth]{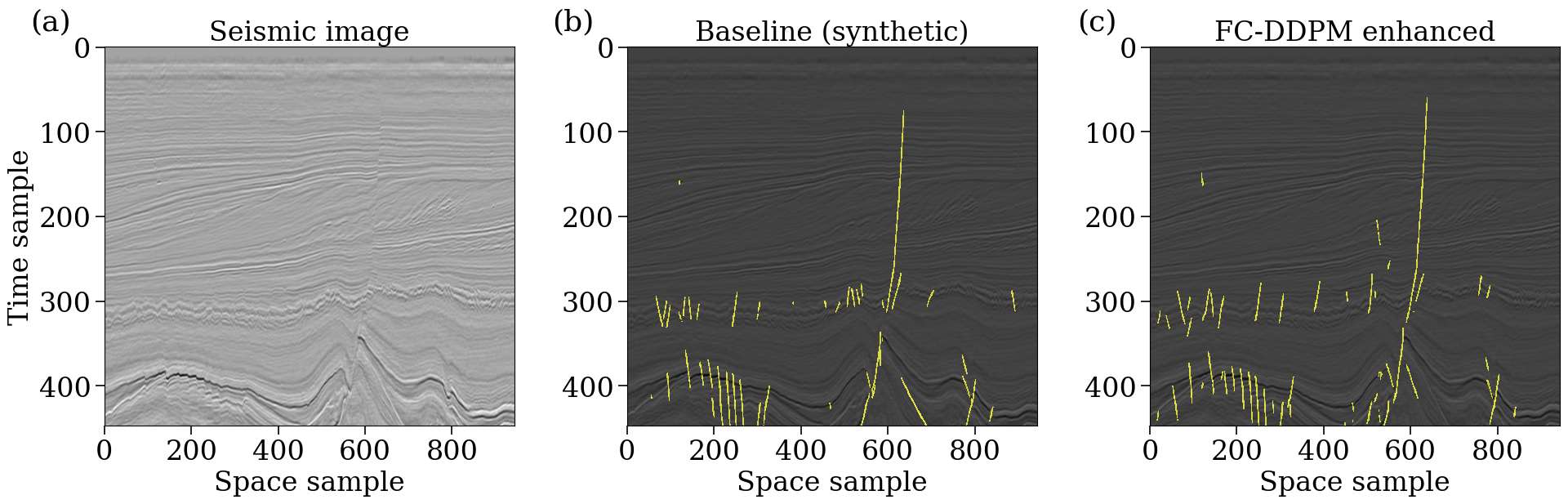}
    \caption{Fault detection results on an inline section from the F3 survey. (a) Original seismic image displaying faulted and folded reflectors in the deeper section (time samples 300–430). (b) Fault predictions from the baseline U-NET trained on 2,000 synthetic samples only. (c) Fault predictions from the same U-Net architecture trained on the 20,000 augmented FC-DDPM-generated dataset. The FC-DDPM-enhanced model yields cleaner predictions with fewer false detections in the unfaulted region and improved continuity along the major fault structures, particularly between crosslines 400 and 800. Both models use identical architecture and training hyperparameters, isolating the effect of data augmentation on prediction quality.}
    \label{fig:Prediction on F3}
\end{figure}

The trained model performs satisfactorily on F3 field data (Figure \ref{fig:Prediction on F3}), further validating the domain relevance of the generated samples. Despite being trained primarily on synthetic and generated data, the network successfully identifies fault structures in real seismic acquisitions. In addition to the improved accuracy, Figure \ref{fig:Prediction on F3} shows predictions from the two models trained only on the synthetic dataset and from theFC-DDPMmodels-enhanced dataset. While both model gave overall satisfactory results in identifying the major faults in the seismic section, the model trained on the enhanced FC-DDPM dataset tends to achieve more continuity in fault detection and can identify more subtle faults.  This controlled experiment reveals that improved results, particularly with respect to fault continuity and the detection of subtle faults, are aided not only by more diverse training samples but also by the FC-DDPM-generated images used for training, which partially capture the textural and noise characteristics of the field data. We also found that the FC-DDPM augmentation improvement scales with the size of labeled datasets. At low labeled dataset, e.g., \(N_{baseline} = 200\) and \(N_{augmented} = 2000\) , FC-DDPM improved field fault detection average precision by almost double (\(PR-AUC\) increased from \(0.15 - 0.22\)). The finding supports our hypothesis that FC-DDPM-generated images can effectively supplement limited labeled datasets for training generalizable fault detection networks.

\begin{table}[t]
\centering
\caption{Quantitative metric assessment of baseline and augmented model on F3 dataset}
\label{tab:downstream_quant_metric}
\begin{tabular}{lcccc}
\hline
\textbf{Model} & \textbf{IoU} & \textbf{Dice} & \textbf{Precision} & \textbf{Recall}\\
\hline
Baseline        & $0.072$ & $0.134$ & $0.541$ & $0.077$ \\
Augmented         & $0.094$ & $0.173$ & $0.579$ & $0.101$ \\
\hline
\end{tabular}
\end{table}

\begin{figure}[tbp]
    \centering
    \includegraphics[width=\textwidth]{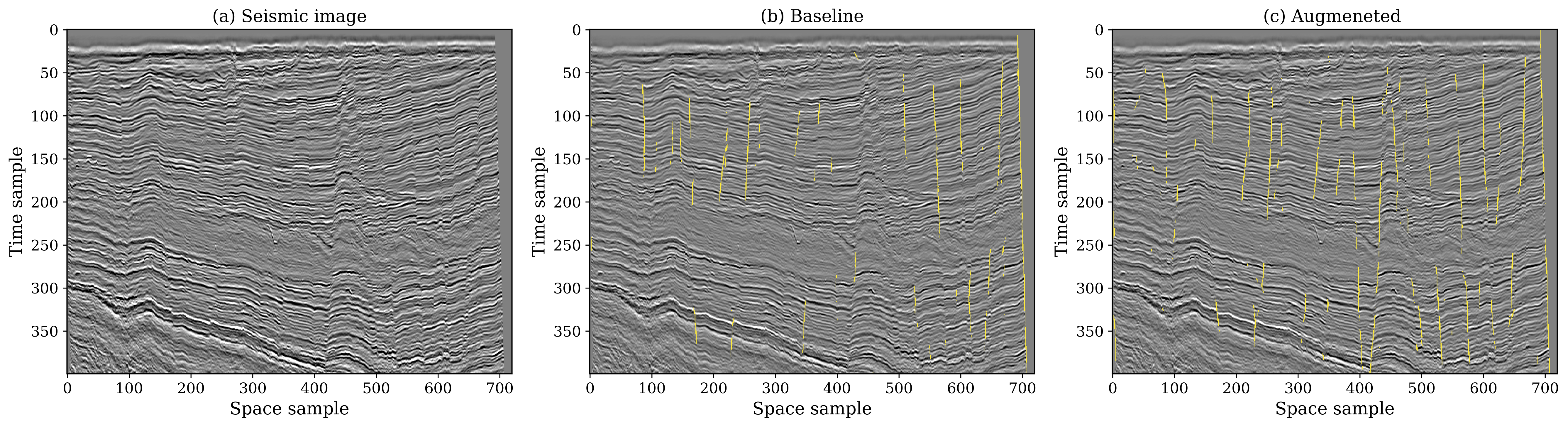}
    \caption{Cross-survey fault detection on the Kerry 3D survey, which was not used during training. (a) A representative inline seismic section. (b) Fault predictions from the baseline model (synthetic only). (c) Fault predictions from the FC-DDPM-augmented model. Both models produce coherent fault detections along the major structural discontinuities of this unseen survey, demonstrating cross-survey applicability. Interpreted fault labels are unavailable for Kerry, so the comparison is qualitative.}
    \label{fig:kerry}
\end{figure}

To assess cross-survey generalization, we applied the trained fault-detection models to the Kerry 3D survey, which was not used in training and for which interpreted fault labels are unavailable (Figure~\ref{fig:kerry}). While results are modest for both models for the obvious discontinuities, the augmented model detects more faults. This complements the quantitative F3 evaluation and reinforces the distinction between in-domain and cross-survey generalization.

\subsection{Broader Implications}
The workflow presented in this study establishes a general template for generating labeled seismic training data when paired annotations are unavailable. While we have focused on fault detection, the same methodology extends naturally to other seismic interpretation tasks, including salt body delineation, horizon tracking, facies classification, etc. Each application requires only the substitution of appropriate conditioning masks during the fine-tuning stage. Beyond seismic interpretation, the transfer learning strategy demonstrated here may benefit other geoscientific domains facing similar data scarcity challenges. Well-log synthesis, seismic-to-well tie generation, and subsurface model population are candidate applications in which labeled data are limited but unlabeled observations are plentiful. The key insight that diffusion models can decouple texture learning from structural conditioning suggests a broadly applicable paradigm for scientific data augmentation.
 
\section{Conclusion}
This study presents FC-DDPM, a fault-conditioned denoising diffusion probabilistic model for generating realistic seismic images with corresponding fault labels. The method addresses a persistent challenge in DL-based seismic interpretation: the scarcity of labeled training data. By combining unconditional pretraining on field seismic data with conditional fine-tuning on stylized synthetic data, FC-DDPM learns to generate images that exhibit both the textural characteristics of field acquisitions and precise adherence to prescribed fault geometries. Three main contributions emerge from this work. First, we demonstrate that transfer learning effectively decouples texture learning from structural conditioning in diffusion models, enabling conditional generation without paired field-label data. The unconditional network trained on field data yields learned seismic priors that accelerate convergence of conditional training at ~3,000 timesteps. Second, although the concept of NST has been proposed and applied to paint field textural characteristics onto synthetic seismic data \cite[]{takemoto2019enriching, jun2022realistic, jiang2023implementation}, we repositioned NST as an intermediate data preprocessing step, enabling its adoption for similar tasks. We show that style transfer serves as an effective domain bridging technique. The F-K spectral analysis confirms that the resulting generated images inherit the spectral properties of field data. Third, we validate the practical utility of the generated data through downstream fault-detection experiments. Augmenting training data with 20,000 FC-DDPM-generated samples reduces overfitting and improves model accuracy and generalization to field data, without architectural modifications.  While the current implementation is limited to DL-based fault detection, the demonstrated workflow can be applied to other DL-based seismic interpretation tasks or to tasks involving unlabeled field and synthetic seismic data, providing a practical path toward more robust seismic-DL tasks.

\FloatBarrier
\bibliographystyle{plainnat}
\bibliography{references}

\end{document}